\documentclass[journal,12pt,onecolumn,draftclsnofoot,]{IEEEtran}

\IEEEoverridecommandlockouts
\def\BibTeX{{\rm B\kern-.05em{\sc i\kern-.025em b}\kern-.08em T\kern-.1667em\lower.7ex\hbox{E}\kern-.125emX}}

\usepackage{settings}
\usepackage{appendix}

\newboolean{short} 
\setboolean{short}{false}
\newboolean{anony} 
\setboolean{anony}{false}

\begin{document}

\title{
Scheduling with Probabilistic Per-Packet Real-Time Guarantee for URLLC
\ifthenelse{\boolean{anony}}
{
\vspace*{-0.2in}
}
{\thanks{This work is supported in part by the NSF awards 2212573, 2229654, 2232461, 2112606, 2130889, and 1827211 and NIFA award 2021-67021-33775.} }
\thanks{Zhibo Meng and Hongwei Zhang are with Iowa State University, U.S.A.  E-mail:\{zhibom,hongwei\}@iastate.edu.}
\thanks{James Gross is with KTH Royal Institute of Technology, Sweden. Email: jamesgr@kth.se.}
}

\author{
\ifthenelse{\boolean{anony}}
{}
{
Zhibo Meng, Hongwei Zhang, 
James Gross 
}

}

\maketitle 

\begin{abstract}
For 
ultra-reliable, low-latency communications (URLLC) applications such as mission-critical industrial control and extended reality (XR), it is important to ensure the communication quality of individual packets. 
    Prior studies have considered Probabilistic Per-packet Real-time Communications (PPRC) guarantees for single-cell, single-channel networks
    but they have not considered real-world complexities such as inter-cell interference in large-scale networks with multiple communication channels and heterogeneous real-time requirements. 
    To fill the gap, we propose a real-time scheduling algorithm based on \emph{local-deadline-partition (LDP)}, and the LDP algorithm 
    ensures PPRC guarantee for large-scale, multi-channel networks with heterogeneous real-time constraints. 
We also address the associated challenge of schedulability test. In particular, we propose the concept of \emph{feasible set}, identify a closed-form sufficient condition for the schedulability of PPRC traffic, and then propose an efficient distributed algorithm for the schedulability test. 
    We numerically study the properties of the LDP algorithm and observe that it significantly improves the network capacity of URLLC, for instance, by a factor of 5-20 as compared with a typical method. 
    Furthermore, the PPRC traffic supportable by the LDP algorithm is significantly higher than that of state-of-the-art comparison schemes.
    This demonstrates the potential of fine-grained scheduling algorithms for URLLC wireless systems regarding interference scenarios.
\end{abstract}

\begin{IEEEkeywords}
URLLC, probabilistic per-packet real-time communications (PPRC) guarantee, large-scale cellular networks.
\end{IEEEkeywords}


\section{Introduction} \label{sec:introduction}

Wireless networks such as 5G-and-beyond cellular systems are increasingly being explored for mission-critical sensing and control applications~\cite{Saad2020:6G-vision}. 
In real-time augmented vision, for instance, such wireless networks could enable the fusion of real-time video streams from spatially distributed cameras to eliminate the line-of-sight constraint of natural human vision, enabling see-through functionality regarding obstacles~\cite{Saad2020:6G-vision,gosain2016,Zhang:pktRT}. 
Furthermore, in industrial automation, 
the significant cost of planning, installation, and maintenance of wired fieldbus solutions have made wireless networks attractive for monitoring and control.
 
Unlike traditional, best-effort wireless networks designed for high-throughput applications, reliable and real-time delivery of individual packets is critical for sensing and control in these 
ultra-reliable, low-latency communications (URLLC) applications. 
For example, in 
Extended Reality (XR) applications and meta-universe services \cite{Saad2020:6G-vision,XR-maintenance}, real-time delivery of each packet enables seamless, naturalistic 3D reconstruction of real-world scenes (e.g., industrial processes), and consecutive packet loss (or long-delay in packet delivery) may well lead to uncomfortable human experience \cite{Saad2020:6G-vision,Zhang:pktRT}. 
In networked industrial 
control, consecutive packet loss may well lead to system instability and cause safety concerns. 
\ifthenelse{\boolean{short}}
{}
{
    In addition, a packet of sample data will be dropped if the packet has not been successfully delivered by the time a new sample data is collected. In this case, the probabilistic guarantee of real-time delivery of each packet enables the modeling of the packet drop process as a random sampling process, thus facilitating the use of random sampling theories to characterize the impact of probabilistic wireless communications on 
    networked control 
    and facilitating the joint design of wireless networking and control  \cite{Zhang:pktRT}. 
}
While on a per stream-level these use cases and requirements generally fall in the realm of URLLC 
as introduced with Release 16 in 5G, in aggregate the aforementioned requirements go far beyond what the current 5G URLLC systems are capable of providing. 
Advanced approaches are required which in particular are capable of \textit{per packet real-time communication guarantees}.

In the literature, different approaches exist towards such guarantees. 
Several recent studies have considered age-of-information minimization \cite{2021aoi,Kuo:minAOI-scheduling,2022aoi,Modiano:minAOI,Hou:minAOIScheduling}. 
\ifthenelse{\boolean{short}}
{Chen et al$.$~\cite{Zhang:pktRT} have proposed an earliest-deadline-first (EDF) scheduling algorithm that ensures probabilistic per-packet real-time guarantee in single-cell, single-channel settings. }
{}
Nevertheless, these approaches do not address the per-packet real-time communication guarantees required by 
\ifthenelse{\boolean{short}}
{heterogeneous URLLC applications in multi-cell, multi-channel networks. } 
{heterogeneous URLLC applications. }
\ifthenelse{\boolean{short}}
{}
{
Chen et al$.$~\cite{Zhang:pktRT} have proposed an earliest-deadline-first (EDF) scheduling algorithm that ensures probabilistic per-packet real-time guarantee in single-cell, single-channel settings. 
However, EDF-based algorithms tend to under-perform in multi-channel settings, while more importantly these approaches do not address the issue of interference in multi-cell deployments.
Indeed, interference can lead to severe implications for URLLC systems jeopardizing any per-packet guarantee~\cite{2022_Ganjalizadeh_URLLC_Interference}. 
}

Addressing interference in URLLC applications is 
\ifthenelse{\boolean{short}}
{also}
{especially}
important for large-scale cellular networks. 
In many envisioned URLLC applications such as those for industrial process control, factory automation, and precision agriculture farming, the network is expected to be deployed across a large area of space to provide URLLC services to a larger number of nodes. 
Thus it is necessary to deploy multiple base stations (BSes) to provide large spatial coverage leading to well-known interference effects.
While interference coordination is the traditional standard solution for such deployments, the effectiveness of interference coordination in URLLC scenarios is little explored to date~\cite{2021_Mahmood_Comm_Letter_URLLC_Interference,2018_Malik_URLLC_Interference}. 
Crucially, existing interference coordination solutions for URLLC 
cannot provide per-packet real-time guarantees.

\paragraph{Contributions.}
In this work, we tackle the challenge of per-packet real-time guarantees under inter-cell interference constraints in URLLC systems from a novel perspective.   
Central to our work are real-time scheduling algorithms with provable characteristics that ensure Probabilistic Per-packet Real-time Communications (PPRC) guarantee in large-scale, multi-cell, and multi-channel settings. 
Given the large scale and the dynamic, uncertain nature of such networks, it is important for the scheduling algorithm to be amenable to distributed implementation without requiring centralized coordination or centralized knowledge of the whole network. 
Given that every network has a limited communication capacity, 
it is also important to be able to decide whether a set of PPRC requests can be supported by the network and the associated scheduling algorithm. 
Thus, there is the need to develop effective schedulability test algorithms that can be deployed in practice. 
We show that corresponding scheduling algorithms can simultaneously provide real-time guarantees while effectively mitigating interference in multi-cell deployments. 
Our main contributions are as follows:
\begin{itemize}
    \item Building upon the idea of deadline partitioning in real-time computing systems, we develop a distributed scheduling algorithm based on local-deadline-partition (LDP). To the best of our knowledge, the LDP scheduling algorithm is the first distributed URLLC 
    scheduling algorithm that ensures PPRC guarantee in large-scale, multi-cell, and multi-channel networks with heterogeneous real-time requirements. 
    
    \item For the schedulability test of PPRC traffic, we propose the concept of \emph{feasible set},  which bridges the real-time computing systems theory and URLLC. We then identify a closed-form sufficient condition for the schedulability of PPRC traffic, and we propose an efficient distributed algorithm for the schedulability test. 

    \item We also identify a necessary condition for the schedulability of PPRC traffic, and use numerical studies to understand a lower bound on the approximation ratio of the LDP algorithm and associated schedulability test. 
    
    \item We numerically study the properties of the LDP algorithm and observe that the intra-cell and inter-cell interference coordination via the LDP scheduling algorithm can significantly increase the capacity of URLLC, for instance, by a factor of 5-20 as compared with the typical industry practice today that only considers intra-cell interference. We also observe that the PPRC traffic supportable by the LDP algorithm is significantly higher than that of the  state-of-the-art algorithms G-schedule \cite{TanWGDH15} and WirelessHART-based algorithm (WH) \cite{2019distributedhart}. For instance, the LDP algorithm can support the PPRC requirement of a large network 32.25\% and 18.41\% of whose links cannot be supported by G-schedule and WH respectively. 
\end{itemize}

The rest of the paper is organized as follows. We summarize related work in Section~\ref{sec:relatedWork}, and we present the system model and problem definition in Section~\ref{sec:preliminaries}. We present, in  Section~\ref{sec:pktRTM}, the LDP real-time scheduling algorithm and the associated schedulability test algorithm. We evaluate the properties of the LDP algorithm in Section~\ref{sec:exptStudy}, and we make concluding remarks in Section~\ref{sec:concludingRemarks}. 
\ifthenelse{\boolean{short}}
{
\emph{Due to the limitation of space, we relegate to the technical report \cite{LDP-TR} the notation summary as well as the proofs of the theorems and lemmas of the paper.}
}
{
}

\section{Related Work} \label{sec:relatedWork}

\ifthenelse{\boolean{short}}
{}
{
\subsection{Real-time computing scheduling}

Real-time scheduling has been well studied in real-time operating systems for 
applications that process data and events with real-time constraints. A scheduling algorithm defines how tasks are processed by the scheduling system, and it  
determines how priorities are assigned to individual tasks. It is not always possible to meet the real-time requirements, and schedulability test and admission control are used to ensure that tasks accepted by a real-time system can be completed before their deadlines. In a uniprocessor real-time system (i.e., only one task can execute at each time slot), the earliest-deadline-first (EDF) scheduling algorithm is optimal for periodic tasks. 
However, EDF is not optimal for multiprocessor systems \cite{2009schedulability}. In a multiprocessor real-time system with $N$ processors (i.e., no more than $N$ tasks can execute at each time slot), p-fair is the optimal solution for periodic tasks 
\cite{1993proportionate}. 
    In this study, we leverage the aforementioned results in real-time scheduling and address unique challenges posed by URLLC in 5G-and-beyond networks. 
}

\ifthenelse{\boolean{short}}
{}
{
\subsection{Real-time communications in wireless networks
}

URLLC has seen significant developments in several areas. Information theory has established the fundamental limits of URLLC capacity, revealing a trade-off between transmission rate and transmission reliability, which depends on the communication channel properties. Stefanovic et al$.$\cite{2017frameless} used information theory to derive reliability-latency guarantees and maximizes reliability within a given latency. 
    Low-latency systems have been designed for URLLC, where numerology is an essential consideration in optimizing wireless communication systems in terms of data rate, latency, and reliability. Wang et al$.$\cite{2021joint} studied the joint design of resource block allocation and beamforming with mixed-numerology for eMBB and URLLC use cases. 
Additionally, Hybrid ARQ is a typical method to increase reliability for URLLC use cases. Nadeem et al$.$\cite{2021nonorthogonal} enhanced the HARQ retransmission mechanism to achieve reliability with guaranteed packet-level latency in the finite block length regime. 
    Karzand et al$.$\cite{2017design} considered the design of FEC tailored for low end-to-end latency requirements in 5G networks, and introduced a new low-delay code construction. Moreover, scheduling is developed at the sender side to achieve high reliability. Zhang et al.\cite{PRK} have identified the Physical-Ratio-K (PRK) interference model, which defines pairwise interference relations between close-by nodes only while ensuring communication reliability.
}

By controlling the allocation of available radio resources (e.g., time, frequency, and power) to meet the specific requirements of URLLC applications, resource allocation is an essential aspect of algorithm development for URLLC. 
    Firstly, URLLC resource allocation has been studies as as an \emph{optimization} problem. For instance, 
\ifthenelse{\boolean{short}}
{}
{
Karimi et al. \cite{karimi2020} proposed a packet scheduling algorithm that reduces queuing delay and improves transmission rate in the downlink, but it is centralized and lacks a method for schedulability test. Anand et al. \cite{anand2018} investigated the minimum system bandwidth required to achieve the necessary URLLC reliability and used one-shot transmission to fulfill latency requirements, but they assumed homogeneous traffic loads. Van et al. \cite{van2022urllc} employed a digital-twin edge wireless network to optimize user association, transmit power, offloading policies and processing rate to meet the latency and reliability requirement, but this approach relies on global information for the entire system. 
}
    Li et al$.$ \cite{li2023joint} aims to optimize bandwidth allocation and power control for both uplink and downlink transmissions, with the objective of minimizing the average total power while ensuring URLLC transmission requirements. However, this approach assumes the availability of global information and homogeneous delay constraints. Wang et al$.$ \cite{wang2023task} focuses on investigating the computational latency and total energy consumption in cell-free massive MIMO systems. These factors are considered as the total cost, and the study examines minimizing this cost by controlling bandwidth and task allocation. However, the approach does not 
guarantee reliability nor latency in the optimization. 
Secondly, \emph{machine learning} methods have also been applied in URLLC resource allocation algorithms. For instance, resource allocation with unsupervised learning in URLLC has been studied in \cite{duong_2023}, where delay and reliability constraints are converted into bandwidth constraints, and unsupervised deep learning is used to find optimal bandwidth allocation.  Yu et al$.$ \cite{yu2023asynchronous} presents a multi-agent reinforcement learning algorithm structure to optimize the overall completion time of uplink and downlink processes, as well as minimizing energy consumption in downlink transmissions. In addition, Min et al$.$ \cite{min2023meta} introduces objective-specific meta-scheduling policies using deep reinforcement learning techniques. The proposed approach considers user utilities, such as average data rate and latency requirements, as the basis for optimizing scheduling decisions. However, these ML methods do not guarantee the latency and reliability requirements of URLLC. 
    Thirdly, \emph{3GPP} includes technology elements for supporting URLLC. Hamidi et al. \cite{hamidi20215g} demonstrated how 3GPP Release 16 extends the real-time capabilities of Release 15 for enhancing latency and reliability; these technology elements alone do not guarantee meeting URLLC requirements, and careful resource allocation methods such as what we present in this paper are needed. The need for URLLC real-time scheduling has also been recognized in recent 3GPP Release 18 standard \cite{XR:standards}. 

Real-time communications have also been extensively studied for industrial wireless networks.
Initial works applied real-time systems scheduling algorithms such as earliest-deadline-first (EDF) and rate-monotonic (RM) to wireless systems.
Chen et al$.$ \cite{Zhang:pktRT} and Wu et al$.$ \cite{WuSGSLC14} have proposed EDF-based scheduling algorithms which ensure per-packet real-time and reliability guarantee. However, both work have avoided wireless channel spatial reuse to satisfy reliability requirements, and the scheduling algorithms therein are not applicable to multi-cell settings. 
Real-time scheduling in WirelessHART industrial wireless networks has been studied \cite{2019distributedhart,Modekurthy2018,Modekurthy2019,Gong2019,SaifullahXLC10,SaifullahGTSLLW15}. They apply EDF, deadline-monotonic, or fixed-priority scheduling algorithms which under-perform in multi-channel settings \cite{Modekurthy2018,Modekurthy2019,SaifullahXLC10,SaifullahGTSLLW15}. In addition, they do not consider channel spatial reuse \cite{Gong2019} as frequently applied in multi-cell deployments.
\ifthenelse{\boolean{short}}
{}
{
In contrast, Xu et al$.$ \cite{XuLWT12} have considered EDF and RM scheduling algorithms and studied the corresponding admission control problems for different interference models in one-channel settings; without using multiple channel available in typical wireless networks, the solutions tend to suffer from limited throughput and do not provide predictable communication reliability. 
Gunatilaka et al$.$ \cite{GunatilakaL18} have proposed a conservative channel spatial reuse method in order to satisfy real-time and reliability requirements. However, the method did not consider the probabilistic nature of wireless communications and cannot ensure communication reliability. 
}

In a second group of works, centralized real-time communication solutions are proposed that explicitly address the differences between traditional real-time systems and wireless real-time communications. 
\ifthenelse{\boolean{short}}
{}
{Chipara et al$.$ \cite{ChiparaLR07} have proposed fixed-priority scheduling algorithms for real-time flows and have studied the corresponding schedulability test problems. They have also considered interference relations between links.}
Destounis et al$.$ \cite{abs-1904-11278} have considered the probabilistic nature of wireless communications and tried to maximize the utility subject to real-time and reliability constraints in communication. However, the study did not consider heterogeneous real-time requirements across links, and the proposed approach is only suitable for single-cell settings. 
    Tan et al$.$ \cite{TanWGDH15} and Peng et al$.$ \cite{peng2021novel} have proposed a centralized scheduling algorithm which is optimal for line networks. 
    But the above centralized solutions are difficult to implement in distributed, multi-cell settings in practice, and they did not consider the deadline requirements of individual packets and did not ensure communication reliability.


A third group of works have considered the provisioning of long-term real-time guarantees for wireless systems. For instance, 
mean delay has been considered in distributed scheduling
\ifthenelse{\boolean{short}}
{\cite{Lin:buffer-sch}}
{
\cite{Lin:buffer-sch,Srikant:elastic-inelastic-traffic-NUM,Jinwoo:provable-csma,Koushik:delay-guarantees}, 
}, 
and age-of-information (AoI) has also been considered in recent studies
\ifthenelse{\boolean{short}}
{\cite{2021aoi}.}
{
\cite{2021aoi,Kuo:minAOI-scheduling,Modiano:minAOI,Hou:minAOIScheduling}.} However, these work have not considered ensuring predictable timeliness of individual packet transmissions in multi-cell, multi-channel network settings. 
\ifthenelse{\boolean{short}}
{}
{
Li et al$.$ \cite{Lou:perNodeAoI} have considered per-node maximum AoI in scheduling, but the study has considered single-cell settings without addressing the challenges of per-node AoI assurance in multi-cell, multi-channel networks. 
}



\ifthenelse{\boolean{short}}
{}
{
}

Table~\ref{table:related work} summarizes the key differences between the LDP scheduling algorithm and existing work. 
\ifthenelse{\boolean{short}}
{
\begin{table}[!htbp]
\small
\centering 
\caption{Related work in real-time scheduling }
\label{table:related work}
\begin{tabular}{|p{2.4in}  |p{.5in} |p{.8in} |p{.7in}| p{.5in} |p{.7in} |}
 \hline
 & Multi-cell wireless network & Predictable per-packet real-time guarantee & Heterogeneous real-time requirement & Distributed algorithm & multi-channel optimal scheduling \\ \hline

 EDF or RM-based scheduling \cite{Zhang:pktRT,2019distributedhart,WuSGSLC14,Modekurthy2018,Modekurthy2019,Gong2019,SaifullahXLC10,SaifullahGTSLLW15} &  & {\LARGE \checkmark } &  {\LARGE \checkmark } &  {\LARGE \checkmark } &\\ \hline
 
Centralized scheduling \cite{TanWGDH15,abs-1904-11278,peng2021novel} &  {\LARGE \checkmark } &  {\LARGE \checkmark } &  &  &\\ \hline
 \ifthenelse{\boolean{short}}
{Mean delay distributed scheduling \cite{Lin:buffer-sch}, age-of-information (AoI) studies \cite{2021aoi} &  {\LARGE \checkmark } &  &  &  {\LARGE \checkmark } &\\ \hline}
{
 Mean delay distributed scheduling\cite{Lin:buffer-sch,Srikant:elastic-inelastic-traffic-NUM,Jinwoo:provable-csma,Koushik:delay-guarantees}, age-of-information (AoI) studies \cite{2021aoi,Kuo:minAOI-scheduling,2022aoi, Modiano:minAOI, Hou:minAOIScheduling} &  {\LARGE \checkmark } &  &  &  {\LARGE \checkmark } &\\ \hline
 }
 Resource allocation for URLLC \cite{li2023joint, wang2023task,duong_2023,yu2023asynchronous,min2023meta,hamidi20215g} & {\LARGE \checkmark }  &  &  &  &{\LARGE \checkmark }\\ \hline
 LDP scheduling &  {\LARGE \checkmark } &  {\LARGE \checkmark } &  {\LARGE \checkmark } &  {\LARGE \checkmark } &  {\LARGE \checkmark }\\ \hline
\end{tabular}
\vspace*{-0.1in}
\end{table}
}
{
\begin{table}[!htbp]
\small
\centering 
\caption{Related work in real-time scheduling }
\label{table:related work}
\begin{tabular}{|p{2.4in}  |p{.5in} |p{.8in} |p{.7in}| p{.5in} |p{.7in} |}
 \hline
 & Multi-cell wireless network & Predictable per-packet real-time guarantee & Heterogeneous real-time requirement & Distributed algorithm & multi-channel optimal scheduling \\ \hline

 EDF or RM-based scheduling  \cite{Zhang:pktRT,2019distributedhart,WuSGSLC14,Modekurthy2018,Modekurthy2019,Gong2019,SaifullahXLC10,SaifullahGTSLLW15} &  & {\LARGE \checkmark } &  {\LARGE \checkmark } &  {\LARGE \checkmark } &\\ \hline
 
Centralized scheduling \cite{TanWGDH15,ChiparaLR07,abs-1904-11278,peng2021novel} &  {\LARGE \checkmark } &  {\LARGE \checkmark } &  &  &\\ \hline
 \ifthenelse{\boolean{short}}
{Mean delay distributed scheduling\cite{Lin:buffer-sch}, age-of-information (AoI) studies \cite{2021aoi} &  {\LARGE \checkmark } &  &  &  {\LARGE \checkmark } &\\ \hline}
{
 Mean delay distributed scheduling \cite{Lin:buffer-sch,Srikant:elastic-inelastic-traffic-NUM,Jinwoo:provable-csma,Koushik:delay-guarantees}, age-of-information (AoI) studies \cite{2021aoi,Kuo:minAOI-scheduling,2022aoi, Modiano:minAOI, Hou:minAOIScheduling} &  {\LARGE \checkmark } &  &  &  {\LARGE \checkmark } &\\ \hline
 }
 Multi-cell real-time scheduling \cite{2007schedulability,2006novel} &  {\LARGE \checkmark } &  {\LARGE \checkmark } &  {\LARGE \checkmark } &  &\\ \hline
Resource allocation for URLLC \cite{karimi2020, anand2018, li2023joint, wang2023task,van2022urllc, hamidi20215g,duong_2023,yu2023asynchronous,min2023meta} & {\LARGE \checkmark }  &  &  &  &{\LARGE \checkmark }\\ \hline
\end{tabular}
\vspace*{-0.1in}
\end{table}

}

\section{System Model and Problem Definition} \label{sec:preliminaries}

\subsection{Network model} \label{subsec:networkModel}

The network consists of $m$ base stations (BSes) and $n$ user equipment (UEs). 
The links between BSes and UEs are called cellular links, and the links between UEs are called device-to-device (D2D) links. 
    The corresponding wireless network can be modeled as a \emph{network graph} $G = (V, E)$, where $V$ is the set of nodes (i.e., the union of the BSes and UEs) and $E$ is the set of wireless links. The edge set $E$ consists of pairs of nodes which are within the communication range of each other. 
The network has access to $N$ non-overlapping frequency channels, denoted by $RB$. Time is slotted and synchronized across the transmitters and receivers. Wireless transmissions are scheduled along frequency and time, with each transmission taking place in a specific frequency channel and time slot. All the time slots are of the same length,\footnote{In 5G URLLC, time slots of different lengths are used, and the length of a time slot may be the multiple of that of the shortest time slot. The PPRC methodology of this paper is readily extensible to the case of varying-length time slots by treating the length of the shortest time slot as the smallest time unit. 
Detailed study of this is worthwhile but beyond the scope of this work (which represents a first step towards understanding fundamental PPRC scheduling issues).
}
and, within a time slot, a transmitter can complete the transmission of one packet.

\subsection{Interference model} 

For PPRC guarantee in URLLC applications, it is important to control interference among concurrent transmissions so that a certain link reliability $p_i$ is guaranteed for
\ifthenelse{\boolean{short}}
{each link $i$.}
{each link $i$.\footnote{Interference control is critical even with interference cancellation \cite{MIMO-interference-cancellation,Li:IAC-general}, beamforming \cite{Tianyi:mmWaveInterferenceControl}, and for mmWave networks \cite{Tianyi:mmWaveInterferenceControl}. }}
For agile adaptation to dynamics and uncertainties, distributed solutions are desired. Existing literature, however, are mostly based on the protocol interference model or the physical SINR interference model, neither of which is a good foundation for developing field-deployable, distributed interference control algorithms \cite{PRKS,PRK}. The protocol model is local and suitable for distributed protocol design, but it is usually inaccurate \cite{Samir:interference-model}; 
the physical model has high-fidelity, but it is non-local and not suitable for distributed protocol design \cite{PRKS,PRK}. 
    To address the challenge, 
    Zhang et al$.$ \cite{PRK} have identified the \emph{Physical-Ratio-K (PRK) interference model that defines pair-wise interference relations between close-by nodes only while ensuring communication reliability (i.e., receiver-side SINR)}. 
In the PRK model \cite{PRK}, a node $C'$ is regarded as not interfering and thus can transmit concurrently with the transmission from another node $S$ to its receiver  $R$ in the same frequency band if and only if $P(C', R) < \frac{P(S, R)}{K_{S, R, T_{S,R}}}$, 
where $P(C', R)$ and $P(S, R)$ is the average strength of signals reaching $R$ from $C'$ and $S$ respectively, $K_{S, R, T_{S,R}}$ 
is the minimum rational 
number 
chosen such that, in the presence of background noise and real-world wireless complexities (e.g., multi-path fading, cumulative interference from all 
concurrent transmitters in the network), the probability for  
$R$ to successfully receive packets 
from $S$ is no less than a minimum link reliability $T_{S,R}$.
\ifthenelse{\boolean{short}}
{}
{As 
shown in Figure~\ref{fig:PRK-model}, the PRK 
\begin{wrapfigure}{r}{.3\linewidth}
	\vspace*{-0.05in}
	\centering
	\includegraphics[width=.98\linewidth]{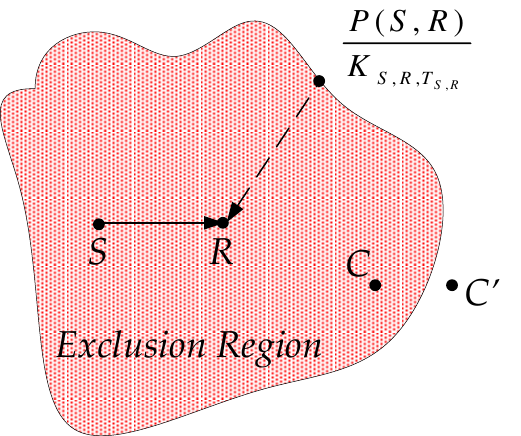}
	\caption{Physical-Ratio-K (PRK) interference model} \label{fig:PRK-model}
	\vspace*{-0.1in}
\end{wrapfigure} 
model defines, for each link $\langle S, R \rangle$, an exclusion region 
around the receiver $R$ such that a node $C$ is in the region 
if and only if $P(C, R) \ge \frac{P(S, R)}{K_{S,R,T_{S,R}}}$. 
Accordingly, every node $C$ in the exclusion region 
is regarded as interfering with and thus shall not transmit concurrently with the transmission from $S$ to $R$ in the same frequency band.

For predictable interference control and to ensure the required communication reliability, the PRK model parameter $K_{S,R,T_{S,R}}$ of each link adapts to the in-situ network and environmental conditions (e.g., node spatial distribution, data traffic load, and wireless channel gain) using a control-theoretic, online learning approach, and, through purely distributed control, the PRK model parameters of all the links quickly converge \cite{PRKS}. 
}
With local, distributed coordination among close-by links alone, PRK-based scheduling achieves 
\emph{close-to-optimal communication capacity} while ensuring the required mean communication reliability \cite{PRKS}. 
The PRK model and PRK-based scheduling have been observed to be \emph{generically applicable} to diverse wireless platforms (e.g., IEEE 802.15.4, LTE, and 5G radios) \cite{PRK,PRKS}, both immobile networks \cite{PRKS} and mobile networks \cite{CPS}, and both ad-hoc networks \cite{PRKS} and cellular networks with D2D links \cite{UCS}, and in real-world scenarios of complex wireless behavior such as multi-path fading and cumulative interference from all concurrent
transmitters in networks.

\begin{wrapfigure}{r}{.5\linewidth}
	\vspace*{-0.00in}
	\centering
	\includegraphics[width=.78\linewidth]{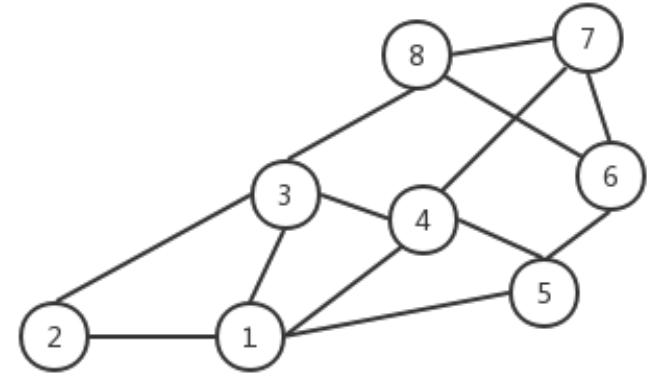}
	\caption{Example conflict graph $G_c$ (Note: this example will be used to illustrate other concepts in the rest of the paper too.)}
        \label{Example}
	\vspace*{-0.03in}
\end{wrapfigure}

To ensure predictable communication reliability, the PRK interference model is used in this 
paper to provide the conflict set information for each link. In particular, a \emph{conflict graph} $G_{c}$ = ($V_c$, $E_c$) is defined for the network $G$, where each node in $V_c$ represents a unique communication link in the network $G$, and $(i, j) \in E_c$ if links $i$ and $j$ interfere with each other, that is, if the transmitter of link $i$ (link $j$) is in the exclusion region of link $j$ (link $i$). Given a link $i$, we let $M_i$ denote the set of links interfering with $i$, that is, $M_i = \{j: (i, j) \in E_c\}$. 
As an example, Figure~\ref{Example} shows a conflict graph with 8 nodes, where each node represents a link in the network $G$. 
Taking link 1 as an example, 
$M_1 = \{2,3,4,5\}$. 

Based on the conflict graph, if one link is active, then none of its interfering links can be active at the same time and frequency. In this way, the mean link reliability of the active links is ensured in the presence of background noise, path loss, fading, and cumulative interference from all concurrent transmitters in the network (including the interference from the links beyond the two-hop neighbors of $i$ in $G_c$). 
In addition, mechanisms such as transmission power control can be used to ensure a certain packet delivery reliability $p_i$ at each time instant based on in-situ network conditions \cite{Zhang:pktReliability}. Based on predictable link reliability enabled by the PRK model, this paper studies how to ensure predictable per-packet real-time communications in multi-cell, multi-channel settings. Therefore, we assume that the link packet delivery reliability for each link $i$ is ensured and denoted by $p_i$.

\subsection{PPRC traffic model}

To support URLLC applications with heterogeneous real-time requirements, we characterize the PPRC data traffic along each link $i$ 
by a 3-tuple  $(T_{i}, D_{i}, S_{i})$:
\begin{itemize}
\item Period $T_{i}$: the transmitter of link $i$ generates one data packet every
$T_{i}$ time slots. 

\item Relative deadline $D_{i}$: each packet along link $i$ is associated with a relative deadline $D_{i}$ 
\ifthenelse{\boolean{short}}
{in units of time slots, and $D_{i} \leq T_i$.}
{
in units of time slots. A packet arriving at time slot $t$ should be successfully delivered no later than time slot $t$ + $D_{i}$; otherwise, the packet is dropped. Since new packets with new information (e.g., sensing data or control signals) are generated every $T_i$ time slots, we assume $D_{i} \leq T_i$. Unlike Chen et al$.$ \cite{Zhang:pktRT}, we don't assume $D_{i} = T_{i}$. Thus both implicit deadlines and constrained deadlines in classic real-time systems are considered. Our model also covers the cases of heterogeneous deadlines across different links.
}
\item PPRC requirement $S_{i}$: due to inherent dynamics and uncertainties in wireless communication, real-time communication guarantees are probabilistic in nature. We adopt the following concept of PPRC 
guarantee first proposed by Chen el al$.$ \cite{Zhang:pktRT}: 
\newtheorem{definition}{Definition}
\begin{definition}
Link $i$ ensures PPRC 
guarantee if $\forall{j}$, $Prob \{ F_{ij}{\leq}D_{i}\}{\ge}S_{i}$, where $F_{ij}$ is the delay (measured in the number of time slots) in successfully delivering the $j$-th packet of link $i$.
\end{definition}
\end{itemize}

For a packet that needs to be successfully delivered across a link $i$ within deadline $D_i$ and in probability no less than $S_{i}$, the requirement can be decomposed into two sub-requirements: 1) successfully delivering the packet in probability no less than $S_{i}$, and 2) the time taken to successfully deliver the packet is no more than $D_i$ if it is successfully delivered \cite{Zhang:pktRT}. 
    Given a specific link reliability $p_i$, the first sub-requirement translates into the required minimum number of \emph{transmission opportunities}, denoted as $X_{i}$, that need to be provided to the transmission of the packet, and $X_{i} = \lceil{\log_{1-p_{i}}{(1-S_{i})}}\rceil$ \cite{Zhang:pktRT}. Note that, when $S_i > p_i$, $X_i > 1$ and retransmissions are required to ensure the required packet delivery probability $S_i$. Then, the second sub-requirement requires that these $X_i$ transmission opportunities are used within deadline $D_i$.\footnote{The $X_{i}$ reserved time slots do not have to be consecutive, and, for real-time packet delivery, they only have to be before the delivery deadline of the packet.}  
Accordingly, the probabilistic real-time delivery requirement for a packet along link $i$ is transformed into a problem of reserving a deterministic number of transmission opportunities, i.e., $X_i$, before the associated relative deadline $D_i$, and $X_i$ is similar to the job execution time in classical real-time scheduling theory. 
    Using $X_i$, we define the \emph{work density} of link $i$ as $\rho_i = \frac{X_i}{D_i}$.

\subsection{PPRC scheduling problem} \label{subsec:PPRC-scheduling-problem}

Based on the aforementioned system model, the PPRC scheduling problem is as follows: Given a network $G = (V, E)$ where each link $i$ has a link reliability $p_{i}$ and PPRC data traffic $(T_{i}, D_{i}, S_{i})$ ($D_{i} \leq T_{i}$), is the set of links schedulable to meet the PPRC 
requirement? If yes, develop an algorithm that schedules the data traffic to satisfy the PPRC 
requirements; 
if not, indicate the infeasibility.

\ifthenelse{\boolean{short}}
{}
{
For ease of reference, Table~\ref{table:notation} summarizes the key notations used in this article. 
\begin{table}[!htbp]
\centering 
\caption{Notation}
\label{table:notation}
\begin{tabular}{|p{.32in} |p{1.15in} |p{.32in}|p{1.15in}|}
 \hline
$T_i$ & The period of link $i$ & $D_{i,j}$ & The absolute deadline of $j$-th packet along link $i$ \\\hline 
$P_i$ & The reliability requirement of link $i$ & $A_{i,j}$ & The arrival time of $j$-th packet along link $i$   \\ \hline 
$p_i$ & The link reliability of link $i$ & $\rho_i$ & The work density of link $i$ \\\hline 
$\sigma_{i,t}$ & The local deadline partition of link $i$ at time $t$ & $d_{i,t}'$ & The beginning time of $\sigma_{i,t}$ \\ \hline 
$d_{i,t}''$ & The absolute deadline of $\sigma_{i,t}$ & $L_{i,t}$ & The length of $\sigma_{i,t}$ \\\hline 
$X_{i}$ & The work demand of link $i$ & $X_{i,t}'$ & The number of times that link $i$ has transmitted for the period at time $t$ \\ \hline 
$X_{i,t}''$ & The remaining work demand of link $i$ at time $t$ & $X_{i,t}$ & The local traffic demand \\\hline 
$\rho_{i,t}$ & The local work density of link $i$ at time $t$ & $M_i$ & The set of conflict links of link $i$\\ \hline 
$K_{i,j}$ & The $j$-th clique of link $i$ that $K_{i,j} \subseteq M_i \cup i$ & $\mathbb{K}_i$ & The set of $K_{i,j}$  \\\hline 
$U_{K_{i,j}}$ & A union of cliques that $K_{i,j} \subseteq U_{K_{i,j}}$ and $U_{K_{i,j}} \subseteq \mathbb{K}_i$ & $S_{i,K_{i,j}}$ & The feasible set that $i \in S_{i,K_{i,j}}$ and $K_{i,j} \in S_{i,K_{i,j}}$ \\ \hline 
$N$ & The number of channels & $mis_{S}$ & A maximal independent set of the set $S$ \\\hline 
$MIS_{S}$ & The set of all maximal independent sets of the set $S$ & $M_{i,2}$ & Two-hop interfering set of link $i$ \\ \hline 
$C_i$ & The maximal conflict set of link $i$ & $c_i$ & A set of the maximal conflict set of link $i$  \\\hline 
$\delta(i)$ & The approximation ratio of link $i$ & $\delta(i)'$ & The topololgy approximation ratio of link $i$ \\ \hline 
\end{tabular}
\vspace*{-0.1in}
\end{table}
}

\section{
URLLC scheduling with probabilistic per-packet real-time guarantee}  \label{sec:pktRTM}

\subsection{Overview} 

For single-channel wireless networks with implicit deadlines (i.e., packet delivery deadlines being equal to inter-packet-generation intervals), Chen et al$.$ \cite{Zhang:pktRT} have shown that an earliest-deadline-first (EDF) scheduling algorithm is optimal for ensuring probabilistic per-packet real-time guarantee. However, just as how EDF scheduling is not optimal in multi-processor systems, EDF-based scheduling is not expected to perform well in multi-channel networks since it cannot support proportionate progress as in fluid models \cite{cho2006}.
\ifthenelse{\boolean{short}}
{Therefore, }
{
For high-performance URLLC scheduling with Probabilistic Per-packet Real-time Communications (PPRC) guarantee in multi-channel 5G-and-beyond networks, therefore, 
}
we turn to optimal multi-processor scheduling for inspiration. In particular, we develop our algorithm based on the idea of deadline partitioning (DP) \cite{cho2006}\cite{levin2010}. 
    In traditional real-time systems, DP is the technique of \emph{partitioning time into slices}, demarcated by the deadlines of all the jobs in the system. Within each slice, all the jobs are allocated a \emph{workload} for the time slice, and these workloads share the same deadline. 
Then, the DP-fair \cite{levin2010} scheduling algorithm allocates a workload to a job in proportion to the \emph{work density} of the job (i.e., the work to be completed divided by the allowable time to complete the work). Therefore, DP-fair ensures proportionate progress in all the jobs and is optimal for computational job scheduling in multi-processor systems.
\ifthenelse{\boolean{short}}
{}
{
Compared with other optimal multi-processor scheduling algorithms such as P-Fair \cite{P-Fair96}, DP-fair allows more freedom in scheduling jobs within a deadline partition. This is because, at each time slot, only the jobs whose local work density is 1 have to be executed and, for the jobs whose work density is less than 1, there is freedom in choosing which one to execute in terms of satisfying the deadline constraints. Such freedom can enable optimizations such as minimizing delay jitter which is an interesting topic for future studies. 
}

Given that the availability of multiple channels (i.e., frequency bands) in 5G-and-beyond 
networks is similar to the availability of multiple processors in multi-processor computer systems, we explore in this study the application of the DP methodology to PPRC scheduling for URLLC applications. 
To this end, we need to address two fundamental differences between multi-cell 5G-and-beyond 
networks and typical multi-processor systems: 
    \emph{Firstly}, not all the links interfere with one another in multi-cell 5G-and-beyond wireless networks, thus each communication channel can be used by more than one link at the same time. 
    Yet the problem of identifying the maximum set of links that can share the same channel is NP-hard itself \cite{PRK}. 
    In addition, even though only close-by links interfere with one another \cite{PRKS} and have to directly coordinate in accessing wireless channels, links far-away from one another are still indirectly coupled due to the chaining effect in connected networks. 
\emph{Secondly}, unlike multi-processor systems where centralized solutions are feasible, dynamic, multi-cell PPRC scheduling requires distributed solutions. 

To address the aforementioned differences and as we will present in detail in Sections~\ref{subsec:LDP} and \ref{subsec:schTest}, we observe that, using the conflict graph to model inter-link interference 
and building upon the multi-channel distributed scheduling algorithm Unified Cellular Scheduling (UCS) \cite{UCS}, the network can be decoupled, and each link only needs to coordinate with the other links in the two-hop neighborhood of the conflict graph in applying DP-based real-time scheduling. Similarly, schedulability can be tested locally at individual links, and the network-wide PPRC traffic is schedulable as long as the link-local schedulability test is positive. 
However, the PPRC scheduling problem is NP-hard as formally shown 
\ifthenelse{\boolean{short}}
{in the online technical report \cite{LDP-TR}.}
{below: 

\begin{theorem} 
    The PPRC scheduling problem is NP-hard. \label{NP-hard}
\end{theorem}
   

        {\begin{proof}
            See Appendix~\ref{appendix:NP-hard}. 
        \end{proof}
        }
}
\noindent Therefore, 
we first develop an approximate solution by extending the traditional DP method  to local-deadline-partition (LDP) real-time scheduling, 
and then we study the associated schedulability test and approximation ratio.

\subsection{Local-deadline-partition (LDP) PPRC scheduling} \label{subsec:LDP}

Each link $i$ and its interfering links in $M_i$ shall not transmit in the same channel at the same time. Thus the set of links in $M_i \cup \{i\}$ can be treated as a conflict set competing for the same set of resources, just as how a set of jobs compete for the same computing resource in a multi-processor system. Therefore, we can extend the concepts of deadline partition, workload, and work density in DP-Fair scheduling \cite{cho2006}\cite{levin2010} to each conflict set.
 In particular, we can define the concepts of local deadline partition, local traffic demand, and local work density to ensure steady, proportionate progress towards completing the required workload (i.e., the number of transmissions required for the PPRC guarantee) within deadlines, and use the local work density to prioritize packet transmissions along different links of a conflict set. 

Unlike traditional real-time systems where the deadline partition (DP) is based on global information (i.e, real-time parameters of all the tasks), the local-deadline-partition (LDP) splits time based only on the information of one-hop links in the conflict graph. 
    In particular, for a link $i \in E$ and $j = 1, 2, \ldots $, let $A_{i,j}$ and $D_{i,j} $ denote the arrival time and absolute deadline of the $j$-th packet along link $i$, respectively; {if the $j$-th packet arrives at the beginning of time slot $t$, $A_{i,j} = t - 1$, and $D_{i,j} = A_{i,j} + D_i = t + D_i -1$. } 
Then, we sort the arrival times and absolute deadlines of the packets along the links in $M_i \cup \{ i \}$ in a non-decreasing order, and regard each non-zero interval between any two consecutive instants of packet arrival/deadline as a \emph{local deadline partition}. More specifically, 
\begin{definition}[Local Deadline Partition] \label{def2}
At a time slot $t$, the \emph{local deadline partition (LDP)} at a link $i \in E$, denoted by $\sigma_{i,t}$, is defined as the time slice $\left[d_{i,t}', d_{i,t}'' \right)$, where $d_{i,t}' = \max\{\max_{k \in M_i \cup \{i\}, D_{k,j} \le t} D_{k,j}, \\\max_{k \in M_i \cup \{i\}, A_{k,j} \le t} A_{k,j}\}$, and 
$d_{i,t}'' = \min\{ \min_{k \in M_i \cup \{i\}, D_{k,j} > t} D_{k,j}, \min_{k \in M_i \cup \{i\}, A_{k,j} > t}  A_{k,j} \}$.

\end{definition}
Note that, different from DP-fair which only uses deadlines for deadline-partition demarcation, LDP uses both arrival times and deadlines in the demarcation. This is because, unlike in traditional real-time computing systems where all the competing jobs are conflicting with one another, not all the links interfere with one another in multi-cell 5G-and-beyond wireless systems. Consequently, as shown in the proof of Lemma~\ref{lemma-3}, using both arrival times and deadlines in deadline-partition demarcation ensures finer-grained proportionate progress in packet transmissions than using deadlines alone in the demarcation, and this finer-grained control of proportionate progress in link packet transmissions is required for ensuring per-packet real-time guarantee in LDP. 

We denote the length of $\sigma_{i,t}$ by $L_{i,t}$, which equals  $d^{''}_{i,t}$ - $d^{'}_{i,t}$.
Let $P_{i,t} = \lceil \frac{t - A_{i,1}}{T_i} \rceil$, then link $i$ is in its $P_{i, t}$-th period at a time slot $t$ for all $t > A_{i,1}$. Let $X_{i,t}'$ denote the number of times that the $P_{i, t}$-th packet at link $i$ has been transmitted along link $i$ till time slot $t$, then $X_{i,t}'' = X_i - X_{i,t}'$ is the remaining work demand of link $i$ at time slot $t$. 
    At the beginning of each deadline partition, we allocate a local traffic demand to link $i$, and it equals the link's remaining work demand multiplied by the ratio of the length of the current deadline partition (i.e., $L_{i,t}$) to the length of the interval between the current time slot and the absolute deadline (i.e., $D_{i,P_{i,t}}-d_{i,t}'$). Inside the deadline partition $\sigma_{i,t}$, the local traffic demand decreases as packets are transmitted in $\sigma_{i,t}$. 
Precisely, we define the local traffic demand and local work density of a local deadline partition $\sigma_{i,t}$ as follows:
\begin{definition}[Local Traffic Demand] \label{def3}
For link $i \in E$ and time slot $t$,
the \emph{local traffic demand} of link $i$ in $\sigma_{i,t}$, denoted by $X_{i,t}$, is as follows:
\begin{equation}
    X_{i,t} = 
    \begin{cases}
     X_{i, d_{i,t}'}'' \frac{L_{i,t}}{D_{i,P_{i,t}}-d_{i,t}'} 
    &  D_{i,P_{i,t}} > d_{i,t}', t =  d_{i,t}' \\
     X_{i, d_{i,t}'} -  (X_{i,t}'-X_{i,d_{i,t}'}')
    & D_{i,P_{i,t}} > d_{i,t}', t > d_{i,t}' \\
    0
    & D_{i,P_{i,t}} \leq d_{i,t}'
    \end{cases}
\end{equation}
where $D_{i,P_{i,t}} \leq d_{i,t}'$ indicates the case of link $i$ having completed its current packet transmissions and thus having a zero local traffic demand at time $t$. 
\end{definition}
\begin{definition}[Local Work Density] \label{def4}
For link $i$, the \emph{local work density} of $\sigma_{i,t}$, denoted by $\rho_{i,t}$, is defined as the ratio of the local traffic demand $X_{i,t}$ to the time duration till the local deadline of completing the transmission of these local traffic. 
That is, $\rho_{i,t} = \frac{X_{i,t}}{L_{i,t}-(t-d_{i,t}')} = \frac{X_{i,t}}{d_{i,t}''-t}$.
\end{definition}
Based on these definitions, we develop the LDP real-time scheduling algorithm by extending the multi-channel distributed scheduling algorithm Unified Cellular Scheduling (UCS) \cite{UCS} to consider PPRC 
requirements.
\begin{algorithm}[!htb] 
\caption{Local-Deadline-Partition (LDP) Real-Time Scheduling at Link $i$ and Time Slot $t$} 
\label{alg1} 
\begin{algorithmic}[1] 
\REQUIRE 
        $A_{i,1}$:  the  arrival  time  of  the first packet along link $i$;\\
        $M_i$: set of interfering links of a link $i \in E$;\\
        $T_l,D_l$: period and relative deadline of link $l \in  M_i \cup \{i\}$:\\
        $X_{i,t}$: local traffic demand at link  $i$;\\
        $State.l.rb.t$: the transmission state of links $l \in M_i \cup \{i\}$ for $\forall rb \in RB$ at time $t$;\\
        $Prio.l.t$: priority of links $l \in M_i \cup \{i\}$;
\ENSURE Perform the following actions at the transmitter and receiver of link $i$: 
\STATE ${State.i.rb.t} = UNDECIDED$, $\forall rb \in RB$; \\
\STATE $Prio.i.t = X_{i,t}/(d_{i,t}''-t)$;\\
\STATE Share $Prio.i.t$ with the links in $M_i$; \\
\STATE done = false;\\
\WHILE{done == false}
    \STATE done = true;\\
        \FOR{each $rb \in RB$ and in the increasing order of the ID of $rb$}
            \IF {having received updates on $State.l.rb.t$ or $Prio.l.t$ from a link $l \in M_i$} 
            \STATE Update the local copy of $State.l.rb.t$ or $Prio.l.t$ at link $i$; 
            \ENDIF 
             \IF{$X_{i,t}$ == 0 \textbf{and} $State.i.rb.t == UNDECIDED$}
            \STATE ${State.i.rb.t}  = INACTIVE$;\\
            \STATE break;
            \ENDIF
            \IF{ $\exists l \in M_i: State.l.rb.t == ACTIVE$}\label{inactive} \label{if inactive}
            \STATE $State.i.rb.t = INACTIVE$;\\\label{inactive}
            \STATE break;
            \ENDIF 
            \IF{$State.i.rb.t == UNDECIDED$ \textbf{and} (($Prio.i.rb.t > Prio.l.rb.t$) \textbf{or} ($Prio.i.rb.t = Prio.l.rb.t$ \textbf{and} ID.i $>$ ID.l)) holds for every UNDECIDED $l \in M_i$}\label{active}
            \STATE $State.i.rb.t = ACTIVE$; \\
            \STATE $X_{i,t} = X_{i,t} - 1$;\\
            \ENDIF 
             \IF{$State.i.rb.t == UNDECIDED$ } \label{if}
            \STATE done = false; \label{done}
            \ENDIF 
\ENDFOR 
\STATE Share $State.i.rb.t$, $\forall rb \in RB$, with the transmitters and receivers of the links in $M_i$; \\
            
\ENDWHILE  
\end{algorithmic} 
\end{algorithm}
In particular, at a time slot $t$ in each local deadline partition, the transmitters and receivers of links set their 
local work densities which are then used to define the links' relative priorities, with links having larger local work densities being given higher priorities in channel access. The transmitter and receiver of a link $i$ compare the priority of link $i$ with its interfering links (i.e., $M_i$) in scheduling, and they execute the following algorithm in a distributed manner: 
\begin{enumerate}[{1)}] 
    \item The transmitter and receiver of each link $i \in E$ initializes its state as UNDECIDED for each channel $rb \in RB$ and calculate its local work density in time $t$. Note that we use the local work density as the priority. Then, the priority will be shared with interfering links through a control channel. 
    \item The transmitter and receiver of link $i$ iterates over the following steps until the state of link $i$ in each channel is either ACTIVE or INACTIVE: 
    \begin{itemize}
    \item For a channel $rb$ in which the state of link $i$ is
    UNDECIDED, if the local traffic demand $X_{i,t}$ is zero or if there exists an interfering ACTIVE link, 
    the state of link $i$ is set as  INACTIVE;
    \item If link $i$ is UNDECIDED and if it has higher priority or the same priority but larger ID than every other UNDECIDED link in $M_i$, the state of $i$ in channel $rb$ is set as ACTIVE, and its local traffic demand $X_{i,t}$ is reduced by one; 
    \item Both the transmitter and receiver of link $i$ share the state of link $i$ with every other node that has at least one associated link interfering with $i$; 
    \item The transmitter and receiver of link $i$ update the state and priority of a link $l \in M_i$, if the transmitter and/or receiver receive a state update about $l$.
    \end{itemize}
\end{enumerate}
If the state of a link $i$ is ACTIVE for channel $rb$ at time slot $t$, link $i$ can transmit a data packet at channel $rb$ and time slot $t$.

The detail of the local-deadline-partition (LDP) scheduling algorithm for time slot $t$ is shown in Algorithm~\ref{alg1}. For conciseness of presentation, the above discussion regards all the links and their associated transmitters and receivers as playing similar roles in executing the LDP algorithm, as in ad hoc networks. For cellular networks, the base stations (BSes) usually take on more roles and assist the user equipment (UEs) in their cells in algorithm execution, and we will briefly discuss this in Section~\ref{subsec:cellularImplementation}.

\begin{wrapfigure}{r}{.65\linewidth}
	\vspace*{-0.3in}
	\centering
	\includegraphics[width=.88\linewidth]{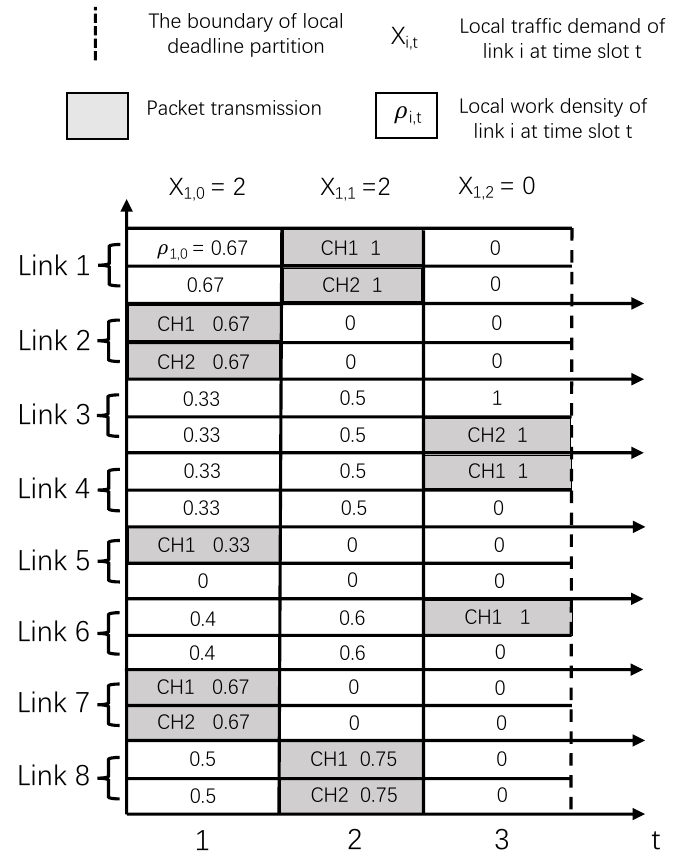}
	\caption{Example of LDP scheduling}
        \label{LDP-example}
	\vspace*{-0.1in}
\end{wrapfigure} 
To illustrate the key concepts of the LDP scheduling algorithm, let's look at how the algorithm is executed for the network whose conflict graph is Figure~\ref{LDP-example}. 
For conciseness of exposition, here we assume the number of channels $N=2$; the key intuition from the example is applicable to general multi-channel settings. 
    Suppose the real-time traffic of a link $i$ is characterized as $\phi_i = (T_{i}, D_{i}, X_{i})$, and the network traffic is such that 
    $\phi_1 = (6,6,4)$, 
    $\phi_2 = (4,3,2)$, 
    $\phi_3 = (6,6,2)$,
    $\phi_4 = (12,12,4)$,
    $\phi_5 = (12,12,4)$,
    $\phi_6 = (6,5,2)$,
    $\phi_7 = (6,6,4)$,
    $\phi_8 = (4,4,2)$.
Then, the scheduling results from time slot 1 to 3 is shown in Figure~\ref{LDP-example}. 
    Let's first focus on link 1. 
    By ordering  the arrival times and absolute deadlines of the packets along the links in $M_1 \cup \{ 1 \}$ in an increasing order, the first local deadline partition for link 1 is  $\left[1,3\right)$ . 
The local traffic demand of link 1 at the beginning of time slot 1 equals the (remaining) traffic demand (i.e., 4) multiplied by the ratio of the length of the deadline partition (i.e., 3) to the duration from the beginning of time slot 1 to the end of the absolute deadline of 6, that is, being  
$4 \times \frac{3}{6-0}=2$, and the priority (i.e., local work density) of link 1 is the local traffic demand divided by the length from the current time slot to the end of the current local deadline partition, that is, $\frac{2}{3}=0.67$.
At the beginning of time slot 1, to decide whether link 1 shall transmit at time slot 1, it compares its priority with those of the links in $M_1$. Even though link 1 has higher priority than links 3, 4, and 5, it has the same priority as link 2 and has a smaller ID than link 2.  Therefore, according to line~\ref{active} in Algorithm~\ref{alg1}, link 2 uses CH1 to transmit a packet and sets its state in channel CH1 as ACTIVE.
Since the ACTIVE link 2 conflicts with links 1 and 3, links 1 and 3 become INACTIVE at time slot 1 for CH1 according to line~\ref{inactive} in algorithm~\ref{alg1} for CH1. 
Following similar analysis, 
\ifthenelse{\boolean{short}}
{Similar analysis can be applied to other time slots and links shown in  Figure~\ref{LDP-example} \cite{LDP-TR}.}
{
links 5 and 7 can be shown to have the highest priority among their interfering links and can be ACTIVE concurrently with link 2 at time slot 1 for CH1. 
    Then, for CH2, link 2 also has the same priority with link 1 and a larger ID than link 1. Therefore, link 2 uses CH2 to transmit a packet and sets its state in channel CH2 as ACTIVE. 
After receiving the state updates from link 2, links 1 and 3 become INACTIVE at time slot 1 for CH1 and CH2 according to line~\ref{inactive} in algorithm~\ref{alg1}. 
    At the beginning of time slot 1, the local traffic demand of link 1 remains 
    2 since no transmission happened for link 1 in time slot 1. The priority of link 1 is 1 since there are only two time slots left for link 1 to complete its remaining transmissions. 
Link 1 has the highest priority among its interfering links in time slot 1, thus link 1 uses both CH1 and CH2 in time slot 1 to complete its transmissions. 
    At the beginning of time slot 2, the priority of link 1 becomes 0 since there is no remaining work demand. 
}

Algorithm~\ref{alg1} can be 
shown to converge for each time slot $t$, and we have 

\begin{theorem} \label{thm-1}
For each frequency channel and time slot, the set of ACTIVE links is a maximal set of links that are mutually non-interfering and have data packets yet to be delivered.
\end{theorem}
\ifthenelse{\boolean{short}}
{
}
{
\begin{proof}
    See Appendix~\ref{appendix:Theorem 2}.
\end{proof}
} 

\subsection{PPRC schedulability test} \label{subsec:schTest}

Given an arbitrary network $G$, it is not always possible to find a schedule to meet the PPRC 
requirement. Therefore, an important task is to determine the schedulability of a set of real-time communication links. To this end, we consider the schedulability of each individual link, and a set of links is schedulable if every link of the set is schedulable. 
Given that a link $i$ interferes with every link in $M_i$, $i$ shares the $N$ wireless channels with the links in $M_i$. Therefore, we try to map the schedulability test of traditional real-time systems into the set of links in $M_i \cup \{i\}$.
    However, unlike tasks in real-time systems where two or more tasks cannot access the same CPU core at the same time, certain links in $M_i$ may not interfere with one another, and those non-interfering links can be active in the same wireless channel and at the same time. For the conflict graph shown in Figure~\ref{Example}, for instance, link 1 can transmit concurrently with link 6, 7, or 8 in the same channel.
Therefore, to utilize the findings from traditional real-time systems, we shall focus on the PPRC traffic demand of the links in every maximal clique $K_{i,j} \subseteq M_i \cup \{i\}$ such that all the links in the clique are interfering with one another.  

For each maximal clique $K_{i,j} \subseteq M_i \cup \{i\}$, $i \in K_{i,j}$, and there could be at most one active link in any channel at any moment in time. 
Due to transmissions along the links other than $M_i \cup \{i\}$, however, it is possible that, for a given wireless channel and time slot, none of the links in a clique $K_{i,j}$ can be active (i.e., when their interfering links are active in the given channel and time slot). For instance, for the conflict graph shown in Figure~\ref{Example}, if links 2, 5, and 8 are active, then none of the links in the clique $\{1,3,4\}$ can be active in the same channel.
    
Therefore, we propose the concept of \emph{feasible set} that, for a given link $i$, jointly considers the PPRC traffic demand of each set of links that is the union of a set of cliques in $M_i \cup \{i\}$ and that, for any given wireless channel and time slot, can have at least one active link in all cases but can have only one active link in the worst case of the transmissions along the links other than $M_i \cup \{i\}$. That is, in a feasible set, there will be at least $N$ number of packets transmitted for each time slot. (As we will show shortly in Theorem~\ref{thm:sufficientCondition}, the concept of feasible set is a foundation for the PPRC schedulability test.) 

More precisely, we define the concepts of \emph{minimum scheduling rate} and \emph{feasible set} to capture the core intuition of this approach. 
\begin{definition}[Minimum Scheduling Rate] \label{def:minSchRate}
Given a conflict graph $G_c$, a set of links $S \subseteq G_c$, and the set of all maximal independent set of $G_c$, denoted by $MIS_{G_c}$, the \emph{minimum scheduling rate of $S$} is $N \times min_{mis \in MIS_{G_c}} |mis \cap S|$, where $|mis \cap S|$ is the number of links in the set $mis \cap S$. 

\end{definition}

\begin{definition}[Feasible Set] \label{def:feasibleSet}
Given a link $i$ and a maximal clique $K_{i,j}$ in the conflict graph $G_c$ such that $i \in K_{i,j}$ and $K_{i,j} \subseteq M_i \cup \{i\}$. Let $\mathbb{K}_i =\{\textrm{maximal clique \ } K_{i,j'}: i \in K_{i,j'} \land K_{i,j'} \subseteq M_i \cup \{i\} \land K_{i,j'} \subseteq G_c \}$, and $U_{K_{i,j}} \subseteq \mathbb{K}_i$ such that $K_{i,j} \in U_{K_{i,j}}$.  
A \emph{feasible set}, denoted by $S_{i,K_{i,j}}$, is defined as the set of links in a $U_{K_{i,j}}$ whose minimum scheduling rate is $N$ (i.e., the number of communication channels in the network). 

\end{definition}

As an example, for the conflict graph shown in Figure~\ref{Example} and the links in $M_1 \cup \{ 1 \}$, there are 3 maximal cliques, that is, $K_{1,1}=\{1,2,3\}$, $K_{1,2}=\{1,3,4\}$, and $K_{1,3}=\{1,4,5\}$. For $K_{1,2}$, the set of feasible sets for $\{1\}$ and $K_{1,2}$, denoted by $\mathbb{S}_{1,K_{1,2}}$, is $\{ \{1,2,3,4 \},  \{1,3,4,5 \},\{1,2,3,4,5 \} \}$. Note that $\{1, 3, 4\}$ is not a feasible set because its minimum scheduling rate is zero, which in turn is due to the fact that $\{2,5,8\}$ is a maximal independent set for the example conflict graph and it does not include any of the links from $\{1, 3, 4\}$. 
    On the other hand, for link $1$ and $K_{1,1}$, the clique $K_{1,1}$ itself is also a feasible set since its minimum scheduling rate is $N$.

The objective of defining the  feasible set concept is to understand the schedulability of PPRC traffic and to enable schedulability test. Therefore, we need to know whether there exists a  feasible set for all the links.
\begin{lemma}
Given a link $i \in E$ and any maximal clique $K_{i,j} \subseteq M_i \cup \{i\}$, there exists at least one  feasible set. 
\label{lemma-1}
\end{lemma}
\ifthenelse{\boolean{short}}
{
}
{
\begin{proof}
See Appendix~\ref{lemma-1}.
\end{proof}
} 

Then, to understand the conditions for schedulability, we first study the conditions under which schedulability is violated. In general, if the work density of link $i$'s interfering links is heavy, then link $i$ is more likely to be unschedulable. Specifically, the violation condition is as follows: 
\begin{lemma}
Given a link $i$ and any maximal clique $K_{i,j}$ such that $i \in K_{i,j}$ and $K_{i,j} \subseteq M_i \cup \{i\}$, if link $i$ misses its absolute deadline at a time slot $t$, then for each feasible set $S_{i,K_{i,j}} \subseteq M_{i} \cup i$, $\sum_{l \in S_{i,K_{i,j}}}\rho_{l,t-1} \ge N+1$.
\label{lemma-2}
\end{lemma}
\ifthenelse{\boolean{short}}
{
}
{
\begin{proof}
See Appendix~\ref{appendix:Lemma2}. 
\end{proof}
} 

Next, we derive a sufficient condition that ensures the schedulability of a link $i$ all the time.
\begin{lemma}
Given a link $i$, if, for every maximal clique $K_{i,j}$ where $i \in K_{i,j}$ and $K_{i,j} \subseteq M_i \cup \{i\}$, there exists a feasible set $S_{i,K_{i,j}}$ such that $i \in S_{i,K_{i,j}}$, $K_{i,j}\subseteq S_{i,K_{i,j}}$, $S_{i,K_{i,j}} \subseteq M_i \cup \{i\}$, and the sum of the work density of all the links in $S_{i,K_{i,j}}$ 
is no more than $N$, then the $X_i$ number of transmissions of each packet at link $i$ will be completed before the associated deadline. 
\label{lemma-3}
\end{lemma}
\ifthenelse{\boolean{short}}
{
}
{
\begin{proof}
See Appendix~\ref{appendix:Lemma3}. 
\end{proof}
} 

Based on Lemma~\ref{lemma-3}, we now derive the schedulability condition. To decide whether a link $i$ is schedulable, we just need to identify the associated feasible set(s) with the minimum sum work density 
and check whether the minimum sum work density is no more than the number of channels $N$. More precisely, the schedulability condition is as follows:
\begin{theorem}[Schedulability Condition] 
Given a link $i$ and the conflict graph $G_c$, let $\mathbb{K}_i$ denote the set of maximal cliques $K_{i,j}$ in $G_c$ such that $i \in K_{i,j}$ and $K_{i,j} \subseteq M_i \cup \{i\}$, and let $\mathbb{S}_{i,K_{i,j}}$ denote the set of  feasible sets for a clique $K_{i,j} \in \mathbb{K}_i$. If $\forall K_{i,j} \in \mathbb{K}_{i}$, we have, 
\begin{equation} \label{sufficientCondition}
    \min_{S_{i,K_{i,j}} \in \mathbb{S}_{i,K_{i,j}}}  \sum_{l \in S_{i,K_{i,j}}} \frac{X_l}{D_l} \le N, 
\end{equation}
then the PPRC traffic of link $i$ can be supported, that is, link $i$ is schedulable. 
\label{thm:sufficientCondition}
\end{theorem}
\ifthenelse{\boolean{short}}
{
}
{
\begin{proof}
See Appendix~\ref{appendix:sufficientCondition}.
\end{proof}
}
From Theorem~\ref{thm:sufficientCondition}, the PPRC schedulability test requires a method of identifying the associated feasible set(s) with the minimum sum work density. 
 To this end, we need an approach to identifying all the feasible sets of interest. By the definition of feasible sets (i.e., Definitions~\ref{def:minSchRate} and \ref{def:feasibleSet}), whether a set $S_{i, K_{i,j}} \subseteq M_i \cup \{i\}$ is a feasible set depends on the maximal independent sets (MIS) of the conflict graph $G_c$. Yet searching for all the MISes of a graph is NP-hard, and, for large graphs, it tends to be computationally undesirable and may even be infeasible in practice. 
    To address the challenge, we observe that, \emph{instead of checking all the MISes of $G_c$, we only need to check the MISes of the subgraph of $G_c$ induced by the links within two-hop distance from link $i$}, since only these links directly impact whether certain links in $M_i \cup \{i\}$ can be active at certain wireless channels and time slots. 
More precisely, we define the Two-Hop Interference Set of a link $i$ and identify two unique properties of feasible sets as follows. 

\begin{definition}[Two-hop Interference Set] \label{def:twoHopIntSet}
Given a conflict graph $G_c$ and a node $i \in G_c$, the \emph{two-hop interference set} of link $i$, denoted by $M_{i,2}$, is the set of links whose distances from $i$ in $G_c$ are two hops.
\end{definition}

Then, we only need to consider $M_{i,2}$ to determine whether a set is a feasible set. For instance, for link $1$ and set $S_{1,K_{1,2}} = \{1,3,4,5\}$ in the example conflict graph and network represented by Figure~\ref{Example}, $M_{1,2} = \{6,7,8\}$, and $M'_1 = \{2, 6, 7, 8\}$. $MIS_{M'_1} = \{ \{2, 6\}, \{2, 7\}, \{2, 8\} \}$. It is easy to verify that, for any of the set $\{2, 6\}$, $\{2, 7\}$, or $\{2, 8\}$, there exists a link in $S_{1,K_{1,2}}$ that does not interfere with any links of the chosen set. Therefore, $S_{1,K_{1,2}}$ is a feasible set. 
    More precisely, we give the following theorem to determine a feasible set.  
\begin{theorem}[Checking Feasible Set]
Given a link $i$, a set of links $S_i$ that is the union of a set of cliques each of which includes $i$ as an element and is a subset of $M_i \cup \{i\}$, define $M'_i = (\{i\} \cup M_i \cup M_{i,2}) \setminus S_i$, and, when $M'_i \neq \emptyset$, denote all the maximal independent sets of $M'_i$ as $MIS_{M'_i}$. 
    When $M'_i = \emptyset$, $S_i$ is a feasible set; when $M'_i \neq \emptyset$, $S_i$ is a feasible set if and only if, for each $mis \in MIS_{M'_i}$, there exists at least one link in $S_i$ that does not interfere with any link in $mis$. 
\label{thm-3}
\end{theorem}
\ifthenelse{\boolean{short}}
{
}
{
\begin{proof}
See Appendix~\ref{appendix:Theorem3}. 
\end{proof}
}

To leverage Theorems~\ref{thm:sufficientCondition} and \ref{thm-3} in developing the PPRC schedulability test algorithm for link $i$, we need a mechanism of identifying the feasible set(s) of minimum sum work density for every maximal clique $K_{i,j}$ in $G_c$ such that $i \in K_{i,j}$ and $K_{i,j} \subseteq M_i \cup \{i\}$. To this end, we observe the following property of feasible sets.
\begin{theorem}[Feasible Set Generation]
Consider a link $i$ and a set of links $S_{i,K_{i,j}}$ such that $i \in S_{i,K_{i,j}}$ and $S_{i,K_{i,j}} \subseteq M_i \cup \{i\}$. For any maximal clique $K_{i,j'}$ in $G_c$ such that $i \in K_{i,j'}$ and $K_{i,j'} \subseteq M_i \cup \{i\}$, if $S_{i,K_{i,j}}$ is a feasible set, then the link set $S_{i,K_{i,j}} \cup K_{i,j'}$ is also a feasible set; if $S_{i,K_{i,j}}$ is not a feasible set, then the set $S_{i,K_{i,j}} \setminus K_{i,j'}$ is not a feasible set either.
\label{theorem:feasibleSetGeneration}
\end{theorem}
\ifthenelse{\boolean{short}}
{
}
{
\begin{proof}
See Appendix~\ref{appendix:feasibleSetGeneration}.

\end{proof}
}

Let $\mathbb{K}_i$ denote the set of maximal cliques $K_{i,j}$ in $G_c$ such that $i \in K_{i,j}$ and $K_{i,j} \subseteq M_i \cup \{i\}$. Then, per Definition~\ref{def:feasibleSet}, for every $K_{i,j} \in \mathbb{K}_i$, every corresponding feasible set $S_{i,K_{i,j}}$ is the links of a subset of $\mathbb{K}_i$ that includes $K_{i,j}$. 
    According to Theorem~\ref{theorem:feasibleSetGeneration}, the feasible set $S^*_{i,K_{i,j}}$ that has the minimum sum work density will be the union of  $K_{i,j}$ and a minimal number of elements in $\mathbb{K}_i \setminus \{ K_{i,j} \}$ that makes a feasible set.
Therefore, for every $K_{i,j} \in \mathbb{K}_i$, to identify the feasible set $S^*_{i,K_{i,j}}$ having the minimum sum work density, we only need to search the subsets of $\mathbb{K}_i \setminus \{ K_{i,j} \}$ of increasing cardinality and stop once every subset of a certain cardinality plus $K_{i,j}$ is a feasible set. Accordingly, we develop Algorithm~\ref{algorithm:schedulabilityTest} for PPRC schedulability test. 

In Algorithm~\ref{algorithm:schedulabilityTest}, we firstly verify if each maximal clique $K_{i,j}$ is a feasible set or not. If it is, then the feasible set with minimum sum work density has been found. Otherwise, let $\mathbb{K}'_i$ denote the set of cliques that cannot be a feasible set individually. This means that for the set of cliques in $\mathbb{K}'_i$, we need to find a combination of cliques that can form a feasible set.
For each $K_{i,j}\in \mathbb{K}'_i$, 
let $A_{c,p}$ denote a union of $c$ number of maximal cliques $K_{i,m} \in \mathbb{K}_i \setminus \{ K_{i,j} \}$, and let $\mathbb{A}_c$ denote the set of all possible $A_{c,p}$. For each $A_{c,p} \in \mathbb{A}_c$, if the set of links in $A_{c,p} \cup K_{i,j}$ is a feasible set according to Theorem~\ref{thm-3}, then the sum work density of the links in $A_{c,p} \cup K_{i,j}$ will be compared with $U_{i,K_{i,j}}$, and the smaller one will be the new $U_{i,K_{i,j}}$. 
    Given a specific set cardinality $c$, if the set of links in $A_{c,p} \cup K_{i,j}$ is a feasible set for each $A_{c,p} \in \mathbb{A}_c$, then the algorithm does not need to check the subsets of greater cardinality and will terminate immediately.

\begin{algorithm} 
\caption{Schedulability Test at Link $i$} 
\label{algorithm:schedulabilityTest} 
\begin{algorithmic}[1]  
\REQUIRE 
        $N$: the number of channels;\\
        $\mathbb{K}_i$: the set of maximal cliques $K_{i,j}$ in $G_c$ such that $i\in K_{i,j}$ and $K_{i,j} \subseteq \mathbb{K}_i$;\\
        $M_{i}$: set of interfering links of a link $i \in E$;\\
        $M_{i,2}$: set of two-hop interference links of a link $i \in E$;\\
        $X_l,D_l$: traffic demand and relative deadline of link $l \in M_i \cup \{i\}$;  \\
\ENSURE whether link $i$ is schedulable;\\
\STATE $U_{i,K_{i,j}} =\infty, \forall K_{i,j} \in \mathbb{K}_i$;
\STATE $\mathbb{K}'_i=\mathbb{K}_i$;
\FOR{each clique $K_{i,j} \in \mathbb{K}_i$} \label{check}
\IF{the set of links in $K_{i.j}$ is a feasible set according to Theorem~\ref{thm-3}, $M_i$, and $M_{i,2}$} \label{check1}
\STATE $U_{i,K_{i,j}} =\sum_{l \in K_{i.j}\} }\frac{X_l}{D_l}$;
\STATE $\mathbb{K}'_i = \mathbb{K}'_i \setminus \{ K_{i,j}$\};
\ENDIF
\ENDFOR

\FOR{each clique $K_{i,j} \in \mathbb{K}'_i$} \label{forclique}
\STATE ${done = 0;}$
\FOR{$c = 1,..., |\mathbb{K}_i \setminus \{ K_{i,j} \} |$} \label{forclique1}
\STATE $done = 1$; 
\FOR{each $A_{c,p}$ (i.e., a union of $c$ cliques from $ \mathbb{K}_i \setminus \{ K_{i,j} \}$
}
\IF{the set of links in $A_{c,p} \cup K_{i.j}$ is a feasible set according to Theorem~\ref{thm-3}, $M_i$, and $M_{i,2}$}
\STATE $U_{i,K_{i,j}} = min(U_{i,K_{i,j}} , \sum_{l \in \{A_{c,p} \cup K_{i.j}\} }\frac{X_l}{D_l})$;
\ELSE
\STATE $done = 0$;
\ENDIF
\ENDFOR
\IF{done == 1}
\STATE $break;$
\ENDIF 
\ENDFOR
\ENDFOR\\ \label{endforclique}

\IF{ $U_{i,K_{i,j}} \le N, \forall K_{i,j} \in \mathbb{K}_i$ 
}
\STATE link $i$ is schedulable;
\ELSE
\STATE link $i$ is not schedulable;
\ENDIF
\end{algorithmic} 
\end{algorithm}

The ``for" loop in line~\ref{forclique} needs  at most $\lvert \mathbb{K}_i \rvert$ iterations, and there will be at most $\lvert \mathbb{K}_i \rvert-1$ iterations for the ``for" loop in line~\ref{forclique1}. 
 For each $c$, there will be ${| \mathbb{K}_i | - 1 \choose c}$ times to check Theorem~\ref{thm-3}, and ${| \mathbb{K}_i | -1 \choose c} \le {| \mathbb{K}_i | -1 \choose \lfloor \frac{| \mathbb{K}_i | -1}{2} \rfloor }$
 for $ \forall c = 1, \ldots, | \mathbb{K}_i| - 1$. 
Therefore, the total computational complexity of Algorithm~\ref{algorithm:schedulabilityTest} is  
$\mathcal{O}({\lvert \mathbb{K}_i \rvert}{(\lvert \mathbb{K}_i \rvert-1)} {| \mathbb{K}_i | -1 \choose \lfloor \frac{| \mathbb{K}_i | -1}{2} \rfloor } )$.
\subsection{Optimality analysis} 

Given that the PPRC scheduling problem (see Section~\ref{subsec:PPRC-scheduling-problem}) 
is NP-hard, the LDP algorithm and the associated schedulability test are approximations of the optimal solutions. 
\ifthenelse{\boolean{short}}
{Here}
{
As a first step towards understanding the optimality of the LDP algorithm and schedulability test, here
}
we develop a necessary condition for PPRC schedulability and use it to derive a lower bound on the approximation ratio of LDP scheduling\footnote{The approximation ratio is defined as, for each link $i$, the PPRC traffic load regarded as schedulable by the LDP algorithm and associated schedulability test (\ref{sufficientCondition}) divided by the PPRC traffic load schedulable by any optimal scheduling algorithm.}.  
\begin{theorem}[Necessary Condition for PPRC Schedulability]
Given a link $i$ and the conflict graph $G_c$, let $\mathbb{K}_i$ denote the set of maximal cliques $K_{i,j}$ in $G_c$ such that $i \in K_{i,j}$ and $K_{i,j} \subseteq M_i \cup \{i\}$. Then, if link $i$ is schedulable, we have 
\begin{equation} \label{necessaryCondition}
	\vspace*{-0.1in}
    \max_{K_{i,j}\in \mathbb{K}_i} \sum_{l \in K_{i,j}}\frac{X_l}{T_l} \le N.
	\vspace*{-0.35in}
\end{equation}
\label{thm-4}
\end{theorem}
\ifthenelse{\boolean{short}}
{
}
{
\vspace{-30pt}
\begin{proof}
See Appendix~\ref{appendix:Theorem6}.
\end{proof}
}

Based on Theorems~\ref{thm:sufficientCondition} and \ref{thm-4}, we can explore the gap between the sufficient condition (\ref{sufficientCondition}) and necessary condition (\ref{necessaryCondition}). In particular, a \emph{lower bound on the approximation ratio}, denoted by $\delta(i)$, is the ratio of the left-hand side of the necessary condition (\ref{necessaryCondition}) to that of the sufficient condition (\ref{sufficientCondition}). That is,
\begin{equation}\label{approximation ratio}
    \delta(i) = \frac{\max_{K_{i,j}\in \mathbb{K}_i} \sum_{l \in K_{i,j}}\frac{X_l}{T_l}}
    {\max_{K_{i,j}\in \mathbb{K}_i}  \min_{S_{i,K_{i,j}} \in \mathbb{S}_{i,K_{i,j}}}  \sum_{l \in S_{i,K_{i,j}}} \frac{X_l}{D_l}}.
\end{equation}


The lower bound depends on two factors: PPRC traffic and network topology, with the former impacting the work densities at individual links and the latter impacting the interference relations among links. 
    Given a specific PPRC traffic, the sum of work density for a set of links increase with the number of links in the set. Hence, to explore the impact of network topology, we define the \emph{topology approximation ratio} as follows. 
    For each clique $K_{i,j}$, $K_{i,j} \in \mathbb{K}_i$, let $S_{min,i,K_{i,j}} = 
        \arg\min_{S_{i,K_{i,j}}\in \mathbb{S}_{i,K_{i,j}}} \sum_{l \in S_{i,K_{i,j}}}\frac{X_l}{D_l}.$
    Then,  the topology approximation ratio can be defined as 
    \begin{equation}\label{topology approximation ratio}
    \delta(i)' = \frac{\lvert K_{max,i,j}'\rvert}{\lvert S_{max,i}\rvert}
    \end{equation}
    where $K_{max,i,j}'$ is the clique in $\mathbb{K}_i$ that has the maximum number of links, 
    and $S_{max,i}$ is, for all $K_{i,j} \in \mathbb{K}_i$, the feasible set $S_{min,i,K_{i,j}}$ that has the maximum number of links.



We can obtain a closed-form solution to the approximation ratio lower-bound for the following network settings: network $G$ is large, the link reliability $p_i$ is the same for all the links, and the exclusive regions of all links include the same number of interfering links. That is, 
 
\begin{theorem}[Approximation ratio lower-bound]
For network $G$, the approximation ratio of algorithm LDP scheduling is greater than $\frac{1}{6}$. 
\label{Approximation ratio bound}
\end{theorem}
\ifthenelse{\boolean{short}}
{
}
{
\begin{proof}
See Appendix~\ref{appendix:Theorem7}.
\end{proof}
}

\ifthenelse{\boolean{short}}
{}
{
As we will show in Section~\ref{sec:exptStudy}, the approximation ratios for typical wireless networks tend to be more than 0.5 and up to over 0.99, demonstrating the close-to-optimal performance of the LDP algorithm and associated schedulability test.  

 \begin{figure*}[!htb]
 \centering
   \begin{minipage}[t]{0.3\linewidth}
     \centering
     \includegraphics[scale=0.1]{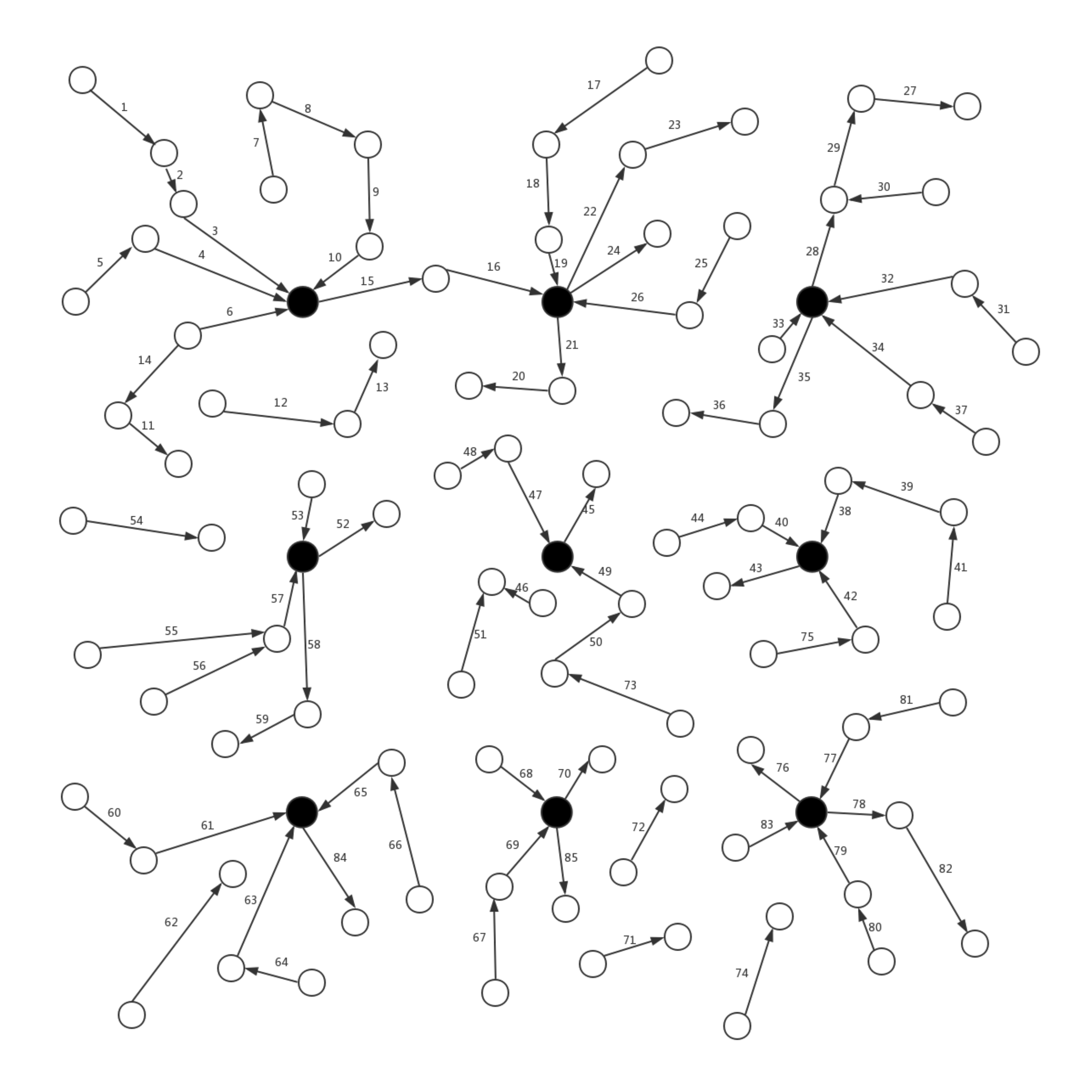}
     \caption{Network 1}
     \label{Fig1}
   \end{minipage}
   \hfill
      \begin{minipage}[t]{0.3\linewidth}
     \centering
     \includegraphics[scale=0.1]{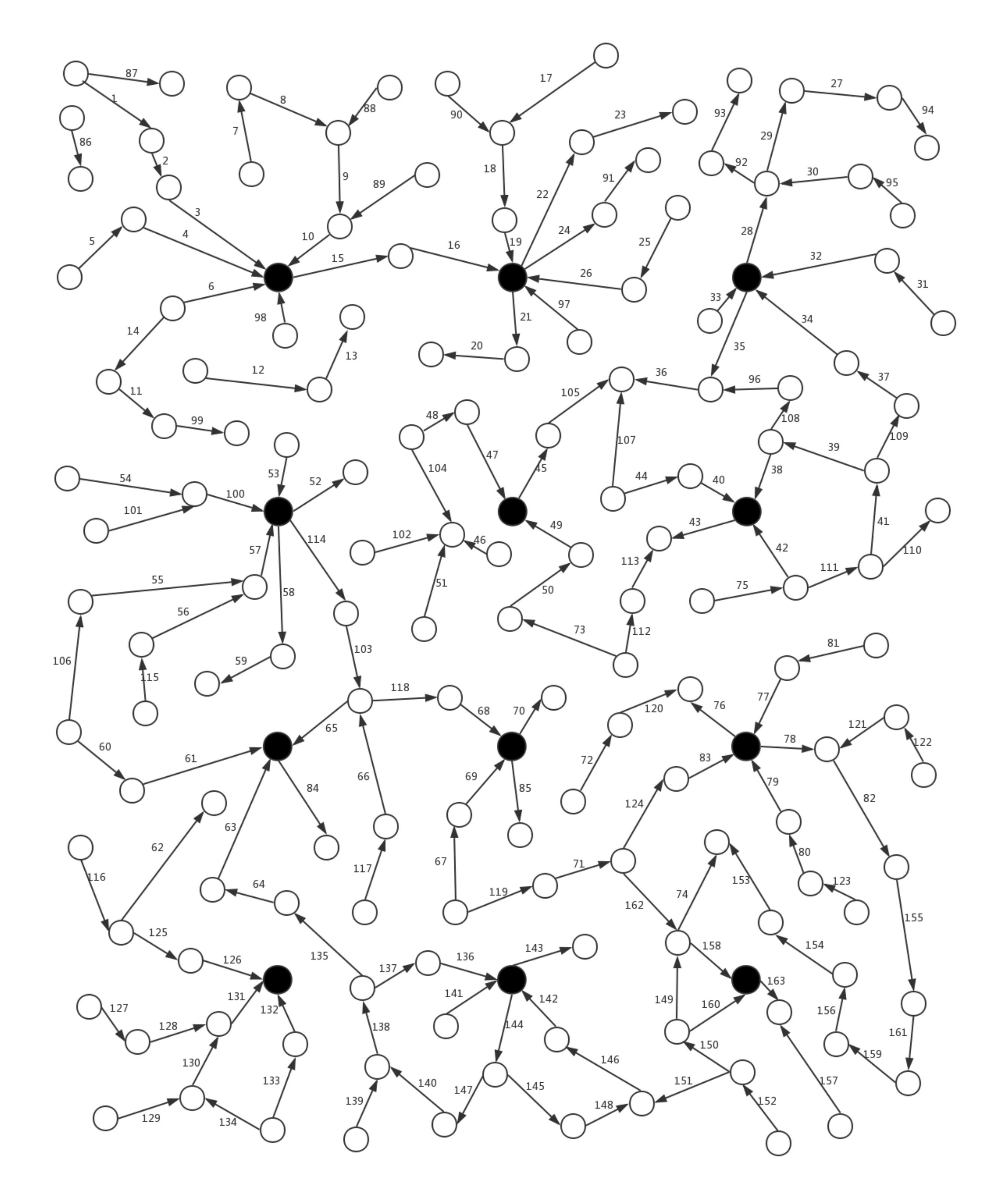}
     \caption{Network 2}
     \label{Fig2}
   \end{minipage}
    \hfill
      \begin{minipage}[t]{0.3\linewidth}
     \centering
     \includegraphics[scale=0.1]{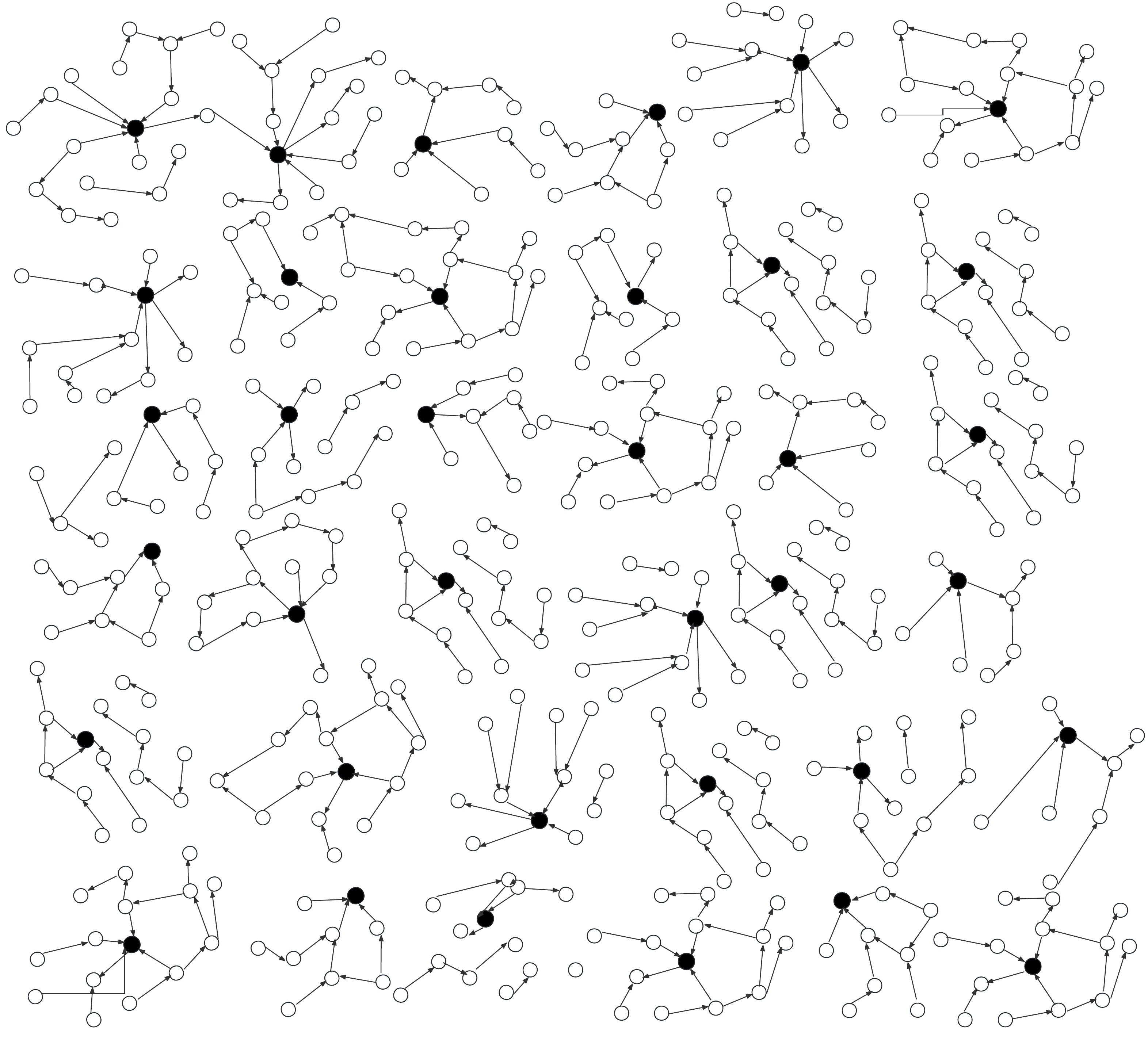}
     \caption{Network 3}
     \label{Fig3}
   \end{minipage}
   
 \end{figure*}
}

\subsection{Remarks: implementation strategies} \label{subsec:cellularImplementation}

While this study mainly focuses on the fundamental algorithmic aspects of the PPRC scheduling problem in URLLC applications, here we briefly present a sketch of an implementation strategy of our approach regarding a cellular network architecture. This strategy builds on the implementation of Unified Cellular Scheduling (UCS) using the open-source cellular software platform OpenAirInterface~\cite{UCS} and USRP software defined radios.
    Both the LDP scheduling algorithm and the PPRC schedulability test algorithm require coordination between base stations (BSes) and user equipment (UEs), as well as among the BSes. The BS-UE coordination can be achieved by using the physical downlink control channel (PDCCH) and physical uplink control channel (PUCCH) to carry relevant control information (e.g., input needed for Algorithm~\ref{alg1}).
Inter-BS coordination can be achieved through the Xn interface, and the Xn interface is usually implemented using high-throughput, low-latency fiber networks.
In terms of quantitative estimates, when it comes to the coordination overhead of the LDP algorithm, a link $i$ only needs to share the priority (1 byte) and state information (2 bits) with other links $l\in M_i$ at each time slot, via their associated BSes. The other necessary inputs can be stored in the BS when the system boots up and only need to be modified if the network changes. 
Based on the information gathered through coordination with its UEs and neighboring BSes, each BS executes the LDP scheduling algorithm 
on behalf of all the cellular and D2D links in its cell. 
Like existing packet transmission scheduling algorithms in cellular networks, the LDP scheduling algorithm is executed at the beginning of each time slot. Given that the input to the LDP algorithm only involves low-overhead, local coordination between a BS and its UEs and neighboring BSes, the control overhead tends to be low and 
is not a barrier to the field-deployment of the LDP algorithm. 
Compared to the LDP algorithm, the PPRC schedulability test algorithm is executed at much lower frequencies and at a timescale of URLLC session dynamics (e.g., emergence of a new URLLC session) and conflict graph dynamics.
Thus, the associated control overhead is even lighter than that of the LDP algorithm.

\section{Numerical Study} \label{sec:exptStudy}

In what follows, we numerically 
evaluate the properties of the LDP scheduling algorithm and the PPRC schedulability test algorithm in diverse multi-cell industrial wireless networks.

\subsection{Network and PPRC traffic settings}

\ifthenelse{\boolean{short}}
{We consider two networks of different sizes.}
{We consider three industrial networks of different sizes and node distribution densities. }
The network size, number of channels, link/node spatial distribution density, and number of conflicting links per link are chosen to represent different real-time network settings, as shown in Table~\ref{table:network settings}.
\begin{table}[!tbp]
\centering 
\begin{tabular}{| p{1.9in} |p{3.3in} |}
 \hline
Communication bandwidth & 20MHz \\\hline 
Number of channels & 3 - 11 \\\hline 
Channel model & Wireless Industrial Indoor path loss model \cite{2016propagation} \\ \hline 
Modulation & 16QAM \\\hline
SINR threshold & 15db \\\hline
Bit error rate & $10^{-6}$ \\\hline
Link reliability & 99\% \\\hline
Packet size & up to 1,000bytes \\\hline 
Network size & 120m $\times$ 120m, 240m $\times$ 240m\\\hline
Number of nodes & 83-320\\\hline
\end{tabular}
\captionof{table}{Network settings}
\label{table:network settings}
\vspace*{-0.05in}
\end{table}
For Network 1, we uniform-randomly deploy 91 wireless nodes in a $120\times120$ square-meter region, generating a network of 83 links. There are nine cells which are organized in a $3\times3$ grid manner. There is a base station (BS) within each cell. 
    For Network 2, we uniform-randomly deploy 151 wireless nodes in a $120\times120$ square-meter region, generating a network of 163 links. There are nine cells which are organized in a $3\times3$ grid manner.
    For Network 3, we uniform-randomly deploy 320 wireless nodes in a $240\times240$ square-meter region, generating a network of 324 links. There are 36 cells which are organized in a $6\times6$ grid manner.
    In addition, we apply the Wireless Industrial Indoor path loss model \cite{2016propagation} to determine the interference effect among links.
\ifthenelse{\boolean{short}}
{
\red{(Detailed topologies of Networks 1, 2, and 3 can be found in \cite{LDP-TR}.) }
}
{}

For Network 1, the maximum and average number of interfering links for a link are 30 and 16.1875 respectively. Network 2 has higher node spatial distribution density and higher degree of cross-link interference, such that the maximum and average number of interfering links for a link are 41 and 23.9438 respectively. Network 3 has larger network scale and similar node density with network 1, and the maximum and average number of interfering links for a link are 33 and 20.713 respectively.
    Regarding the number of channels, with a numerology similar to 5G Numerology 4, the subcarrier spacing is 240KHz, and, assuming that each resource-block (RB) consists of 12 subcarriers, each RB occupies 2.8MHz spectrum. Assuming a communication bandwidth of 20MHz, it gives 7 RBs  
    (i.e., N =7). To represent URLLC scenarios having diverse timing requirements and thus diverse transmission-time-intervals (TTI) and numerologies, the number of channels considered here ranges from 3 to 11.

Assuming that the packet size is 1,000 bytes\footnote{The packet size for URLLC control message may have short packet size, while URLLC media data require large packet size. If the packet size for control messsage is 100 bytes, then the link reliability can achieve 99.9\% according to the network settings, and the required transmission opportunities is at least 4, which is considered in our traffic demand. } 
and 16QAM modulation is applied, the bit error rate could achieve $10^{-6}$ when the SINR threshold is 15db, and the link reliability can achieve 99\%. When the per-packet communication reliability is 99$\%$, we need at least 5 transmission opportunities if URLLC applications require the probability of packet loss or deadline violation to be no more than $10^{-3}$ or even $10^{-9}$.
To experiment with different work densities and to include scenarios of both light and heavy PPRC traffic, the traffic demand $X_i$ (i.e., required number of transmission opportunities per packet) along a link $i$ is uniform-randomly chosen from $[2,5]$.
 Most URLLC use cases such as XR
 can accept 10-20ms one-way delay, thus we assume that the relative deadline $D_i$ uniform-randomly ranges from 6 to 18 time slots.\footnote{Instead of using absolute time values such as 1ms, here we use time-slot as the unit of time specification. Depending on the numerology used in a cellular network, the duration of a time-slot can be configured as 1ms, 0.5ms, 0.1ms etc.} 
    The period is assumed to be greater than or equal to the relative deadline, and we experiment with different periods that differ from the relative deadline by a value uniformly distributed in  $[0,D_i/6]$. 

\subsection{Numerical results}

\paragraph{Approximation ratio of LDP}

\begin{figure*}[b]
  \begin{minipage}[t]{0.3\linewidth}
    \centering
    \includegraphics[scale=0.43]{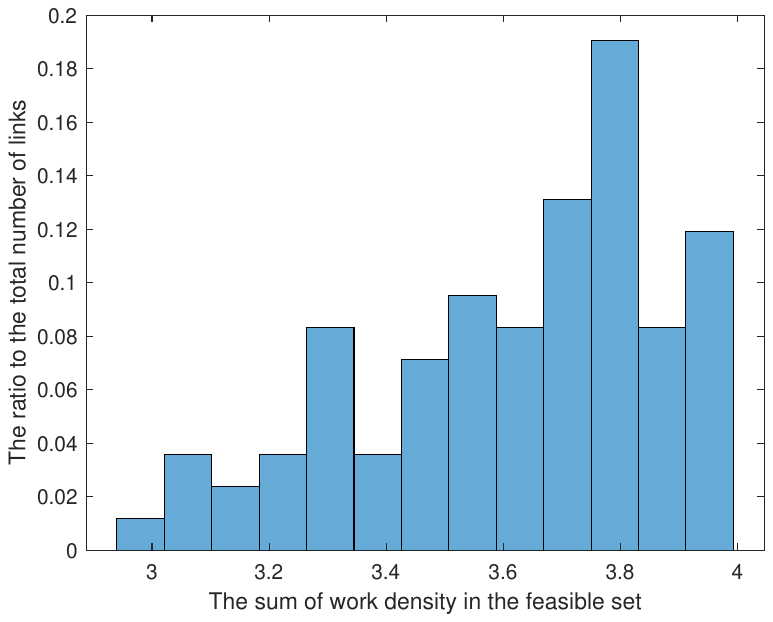}
    \caption{Minimum sum work density of feasible sets}
    \label{sum work density}
  \end{minipage}
   \hfill
    \begin{minipage}[t]{0.3\linewidth}
    \centering
    \includegraphics[scale=0.43]{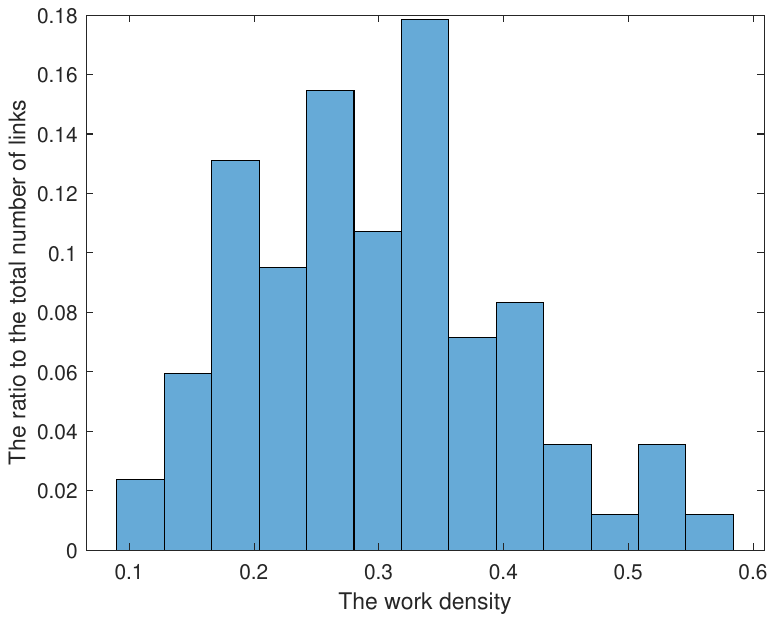}
    \caption{The work density of each link}
    \label{Fig6.sub.2}
  \end{minipage}%
   \hfill
  \begin{minipage}[t]{0.3\linewidth}
    \centering
    \includegraphics[scale=0.43]{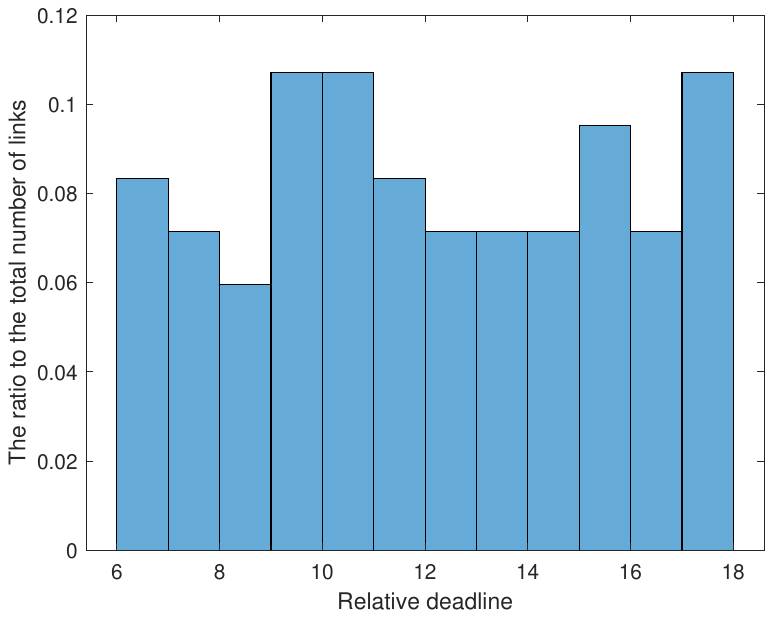}
    \caption{Relative deadline}
    \label{Fig6.sub.1}
  \end{minipage}
\vspace*{-0.15in}
\end{figure*}

\begin{figure*}
  \begin{minipage}[t]{0.31\linewidth}
    \centering
    \includegraphics[scale=0.42]{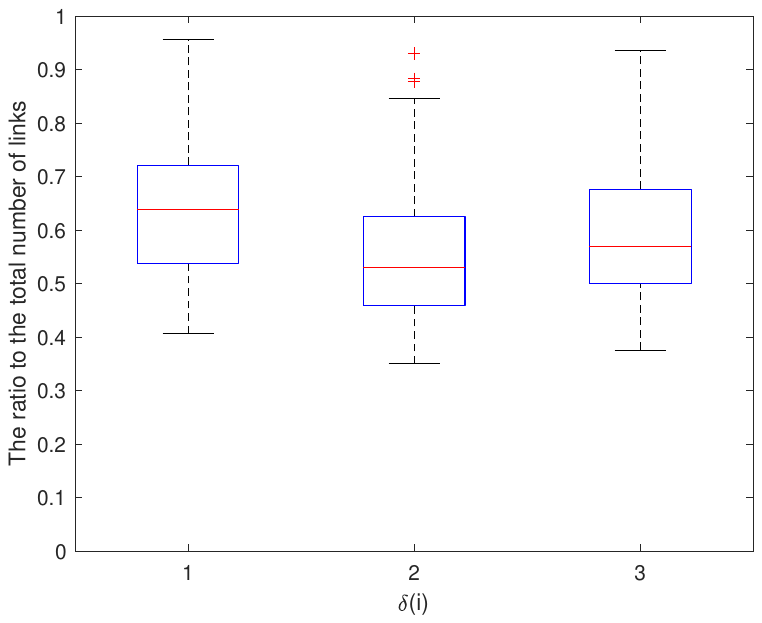}
    \caption{Approximation ratio lower bound}
    \label{Fig3.sub.1}
  \end{minipage}%
   \hfill
  \begin{minipage}[t]{0.3\linewidth}
    \centering
    \includegraphics[scale=0.42]{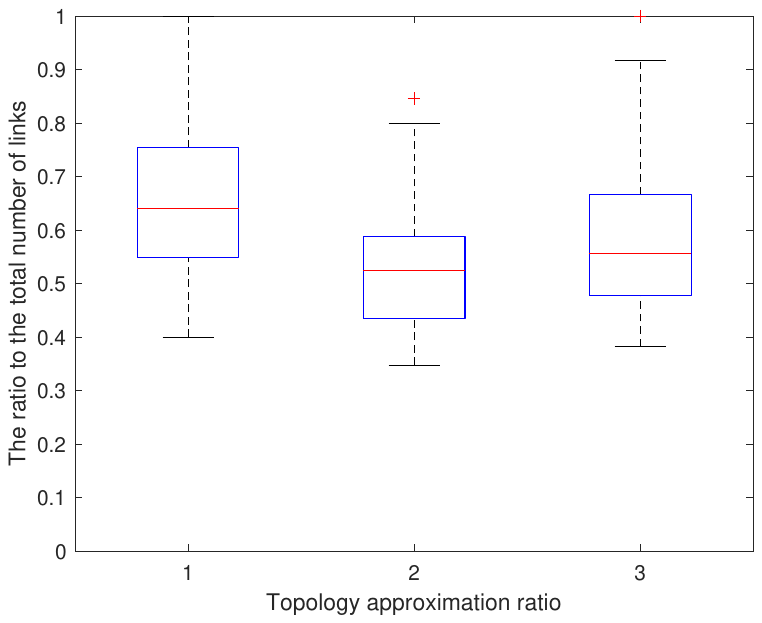}
    \caption{Topology approximation ratio}
    \label{Fig3.sub.2}
  \end{minipage}
   \hfill
  \begin{minipage}[t]{0.3\linewidth}
    \centering
    \includegraphics[scale=0.46]{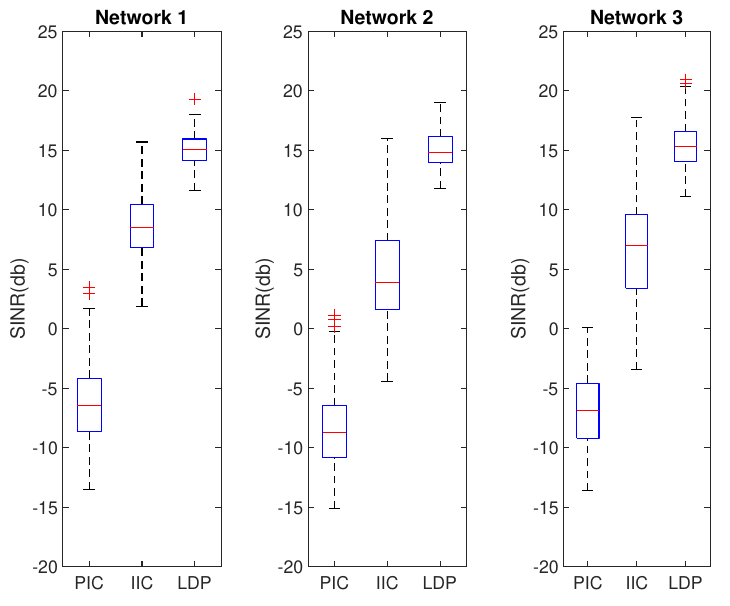}
    \caption{Interference effect on receiver-side SINR}
        \label{interference}
  \end{minipage}
\vspace*{-0.1in}
\end{figure*}

Here we evaluate the lower bound on the approximation ratio of the LDP scheduling algorithm for Networks 1, 2 and 3. To this end, we consider scenarios of demanding PPRC traffic that is close to the network capacity but can still be supported by the LDP algorithm. 
    We take Network 1 as an example, Figure \ref{sum work density}
    shows the histogram of the minimum sum work density of the feasible sets when the number of channels is 4, Figure~\ref{Fig6.sub.2} shows the histogram of the links' work densities, and Figure~\ref{Fig6.sub.1} shows the histogram of the relative deadlines. 
Then, we show the numerical results for the three networks. Figures~\ref{Fig3.sub.1} is drawn from Equation~\ref{approximation ratio}, and they show the approximation ratio lower bound $\delta(i)$ for all the links in Networks 1, 2 and 3 respectively. 
    Figures~\ref{Fig3.sub.2} is drawn from Equation~\ref{topology approximation ratio}, and they show the topology approximation ratio in Networks 1, 2 and 3 respectively. 
For Network 1, the mean approximation ratio lower bound is 0.639, and its 25\%-75\% percentiles is $\left[0.5376, 0.7212\right]$; the mean topology approximation ratio is 0.643, and its 25\%-75\% percentiles is $\left[0.5486, 0.7543\right]$. We see that network topology has significant impact on the approximation ratio, even though the PPRC traffic pattern also impacts the approximation ratio. 
For Network 2, the mean approximation ratio lower bound is 0.5309, and its 25\%-75\% percentiles is $\left[0.4615, 0.6254\right]$; the mean topology approximation ratio is 0.5238, and its 25\%-75\% percentiles is $\left[0.4348, 0.5882\right]$.
For Network 3, the mean approximation ratio lower bound is 0.5691, and its 25\%-75\% percentiles is $\left[0.5006, 0.6767\right]$; the mean topology approximation ratio is 0.5556, and its 25\%-75\% percentiles is $\left[0.47826, 0.6667\right]$. 
    We see that the approximation ratio lower bound in Network 2 and 3 is about 10\% lower than that in Network 1. This is because the number of interfering links per link in Networks 2 and 3  tends to be higher than that in Network 1. Accordingly, the size of cliques in the conflict graph of Networks 2 and 3 is greater than that of Network 1, which makes the approximation ratio lower bound relatively lower in Networks 2 and 3. 
The mean approximation ratio lower bound is more than 0.53 for the three networks, and it is up to 0.9553, 0.933 and 0.941 in Networks 1, 2 and 3 respectively. 
\ifthenelse{\boolean{short}}
{}
{
The approximation ratio lower bounds presented above are the lower bound on the performance of the LDP scheduling algorithm. How to potentially tighten the lower bound to precisely characterize the benefits of using the LDP algorithm will be an interesting topic for future studies.
}

\paragraph{Impact of interference coordination}
To understand the importance of considering interference control in URLLC, we consider the impact of three different interference coordination methods. The first method only considers primary interference control (PIC).  That is, only those links sharing a common transmitter or receiver are regarded as conflicting with one another. 
    The second method only considers primary interference control and intra-cell interference control (IIC). That is, the links in the same cell cannot transmit at the same time slot and through the same frequency channel. The third method
    considers the PRK-based intra-cell and inter-cell interference control which is utilized by LDP, and we set the SINR threshold as 15db. 
Each interference coordination method has its associated conflict graph for Networks 1, 2 and 3 respectively, and we use the LDP scheduling algorithm with the different interference coordination methods to understand their impact. 
For each network, we generate the traffic demand that is close to the respective network capacity but can still be supported by the LDP algorithm, and then measure the SINR value as shown in Figure~\ref{interference}.
For network 1, the mean SINR of PIC is -6.4281db, and its 25\%-75\% percentiles is [-8.6693, -4.1557]; the mean SINR of IIC is 8.5314db, and its 25\%-75\% percentiles is [6.8664, 10.4736];  the mean SINR of LDP is 15.0872db, and its 25\%-75\% percentiles is [14.1316, 15.9487]. 
    For network 2, the mean SINR of PIC is -8.7245db, and its 25\%-75\% percentiles is [-10.8384, -6.4589]; the mean SINR of IIC is 3.8573db, and its 25\%-75\% percentiles is [1.6417, 7.4089]; the mean SINR of LDP is 14.7879db, and its 25\%-75\% percentiles is [13.937, 16.1223]. 
    For network 3, the mean SINR of PIC is -6.9126db, and its 25\%-75\% percentiles is [-9.2237, -4.85]; the mean SINR of IIC is 6.9621db, and its 25\%-75\% percentiles is [3.4166, 9.5726]; the mean SINR of LDP is 15.3074db, and its 25\%-75\% percentiles is [14.0377, 16.5858]. 
Therefore, considering the intra-cell interference and inter-cell interference in LDP ensures the required receiver-side SINR, and it significantly increases the receiver-side SINR as compared with PIC and IIC, e.g., by a margin of over 20db. 

\begin{figure*}[b]
  \begin{minipage}[t]{0.32\linewidth}
    \centering
    \includegraphics[scale=0.31]{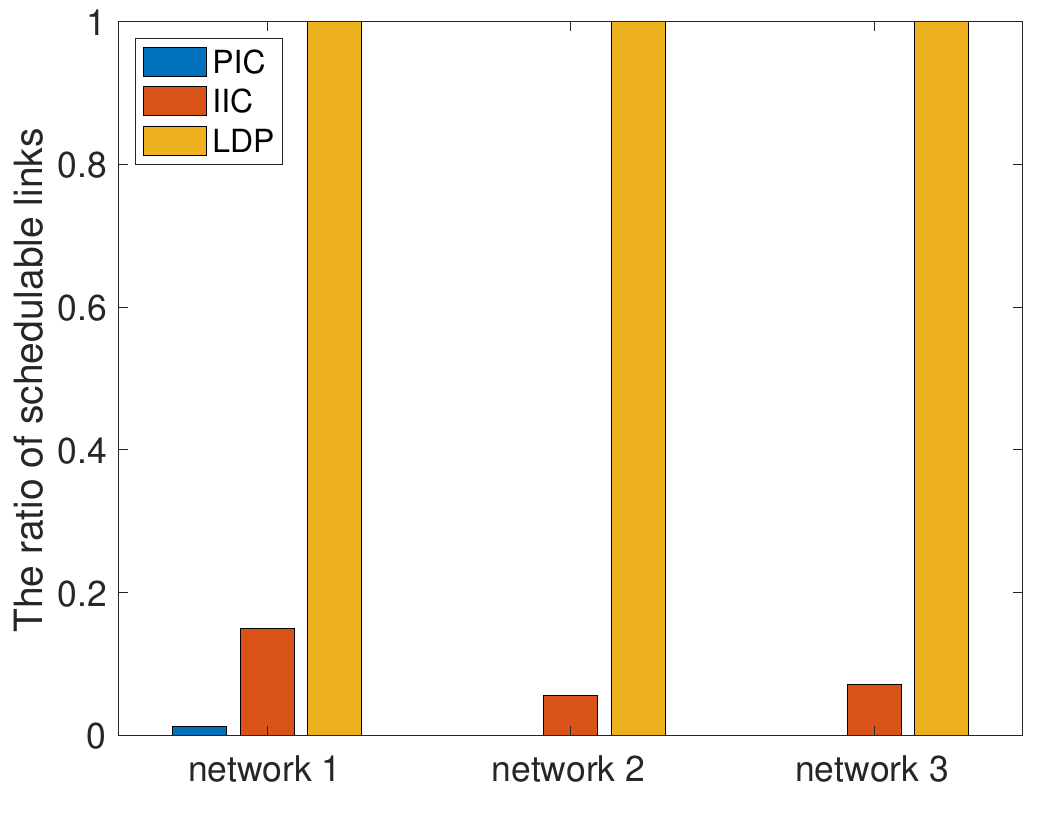}
    \caption{Interference effect on schedulability ratio}
    \label{sr}
  \end{minipage}%
   \hfill
  \begin{minipage}[t]{0.32\linewidth}
    \centering
    \includegraphics[scale=0.31]{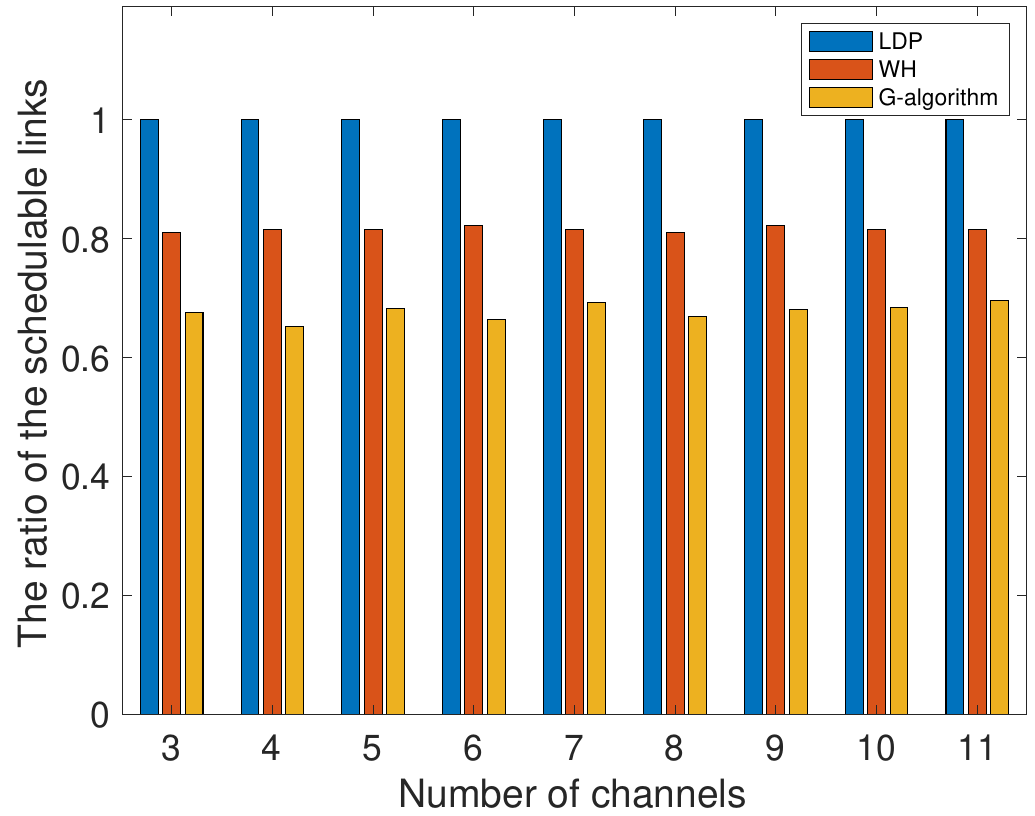}
    \caption{Comparison with G-Schedule}
    \label{Fig:G-algo}
  \end{minipage}
   \hfill
  \begin{minipage}[t]{0.32\linewidth}
    \centering
    \includegraphics[scale=0.31]{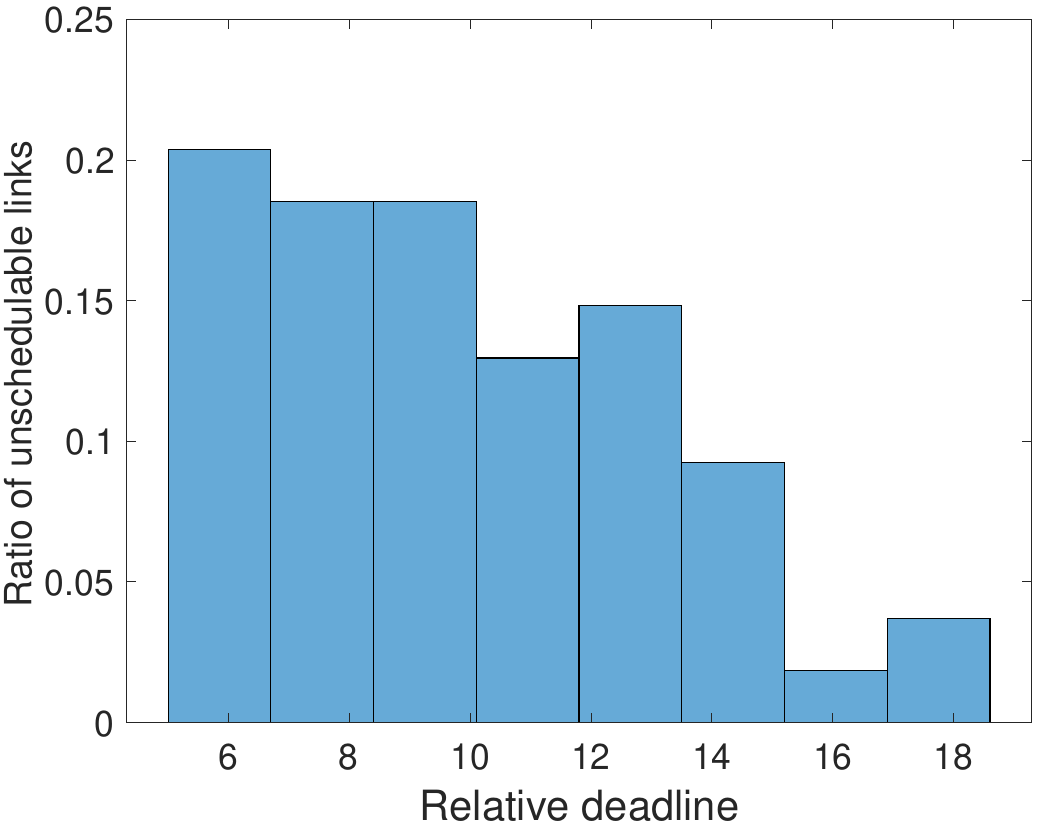}
    \caption{Infeasible links for G-Schedule}
    \label{Fig:Infeasible links}
  \end{minipage}
\end{figure*}
We also consider the impact of the receiver-side SINR on real-time schedulability, which is shown as the ratio of schedulable links in Figure~\ref{sr}. The ratio of schedulable links greatly increases with the decreasing interference. In particular, the ratio of schedulable links of network 1 increase from 0.0125 of PIC, 0.15 of IIC to 1 of LDP; the ratio of schedulable links of network 2 increase from 0 of PIC, 0.05625 of IIC to 1 of LDP; the ratio of schedulable links of network 3 increase from 0 of PIC, 0.0709 of IIC to 1 of LDP. 
    We see that using the LDP scheduling algorithm to address intra-cell and inter-cell interference can significantly increase the real-time capacity (i.e., ratio of scheduable links) by a huge margin, e.g., a factor of about 5-20 as compared with IIC. 
Having demonstrated the impact of considering interference control in URLLC, we next examine the benefit of LDP as compared with other real-time wireless scheduling algorithms that consider interference control. 


\paragraph{Comparative study} 

Out of the existing real-time wireless scheduling algorithms that consider intra-cell and inter-cell interference, the WirelessHART-based algorithm (WH) \cite{2019distributedhart} and G-schedule algorithm \cite{TanWGDH15} address problems that are closest to the PPRC scheduling problem. For real-time multi-channel scheduling in multi-cell cellular networks, the WirelessHART-based algorithm considers the scheduling methods EDF and DM (where the link with the shortest deadline acquires the highest priority), and it gives the worst-case delay analysis and a closed-form schedulability test. 
    G-schedule greedily schedules non-interfering links based on their IDs, and it has been shown to be optimal for the special line networks where all the nodes are located along a straight line \cite{TanWGDH15}. 
To study the performance of LDP and the two other algorithms, we calculate the schedulability test of WirelessHART-based algorithm for each node in Network 2 and implement the LDP and G-algorithm in Matlab and study their behavior in Network 2. (Similar phenomena have been observed for other networks.) 
    We execute each algorithm for 200,000 time slots and observe the ratio of the number of schedulable links (i.e., the links whose probabilistic per-packet real-time requirement is met) to the total number of links. 
\ifthenelse{\boolean{short}}
{}
{
We consider scenarios of demanding PPRC traffic that is close to the network capacity but can still be supported by the LDP algorithm. Then we characterize the feasibility of supporting the real-time traffic using the WirelessHART-based algorithm and G-schedule algorithm. 
}

Figure~\ref{Fig:G-algo} shows the 
ratio of schedulable links in the network. 
We see that, while LDP is able to schedule demanding PPRC traffic (i.e., the ratio of the schedulable links is 100\%), the average ratio of schedulable links in WH algorithm and G-Schedule are 0.8159 and 0.6775. 
    The cause for the difference between WH algorithm and LDP is that the former only considers the worst case for each node and underestimates the feasibility, while LDP can improve such worst case analysis and calculate the schedulability test based on the topology information. The ratio of schedulable links in WH does not increase with the number of channels since the WH-based schedulability test considers the sum work density of a set of links and not the number of channels. 
To understand the cause for the difference between G-Schedule and LDP, we divide the links into different groups according to their relative deadlines, and calculate the ratio of the number of unschedulable links in G-schedule to the total number of links in the corresponding group. 
    Figure~\ref{Fig:Infeasible links} shows the relationship between unschedulable links and their relative deadline. 
We see that the links with shorter deadlines are more likely to become unschedulable in G-schedule.
    This is because G-schedule greedily schedules links 
    without considering heterogeneous deadline constraints, and the links with shorter deadlines tend to be assigned with fewer transmission opportunities with respect to their deadlines. 
On the other hand, LDP dynamically updates packets' priorities based on in-situ work densities, and the links with higher work densities and closer to their absolute deadlines tend to get higher priorities. Accordingly, LDP can support more demanding real-time traffic than what G-schedule can.

\section{Concluding Remarks} \label{sec:concludingRemarks}

For supporting heterogeneous URLLC applications in large-scale 5G-and-beyond networks, we have proposed a distributed local-deadline-partition (LDP) scheduling algorithm to ensure Probabilistic Per-packet Real-time Communications (PPRC) guarantee in large-scale, multi-cell, and multi-channel network settings. The LDP algorithm effectively leverages the two-hop information in the conflict graph and addresses the challenges of multi-cell, multi-channel PPRC scheduling. The concept of feasible set in this paper bridges real-time computing systems and URLLC. 
Leveraging the feasible set concept, we have identified a closed-form sufficient condition for PPRC schedulability test; 
we have also developed an algorithm for finding the minimum sum work density of feasible sets, upon which we have developed the schedulability test algorithm. 
    Our numerical 
    results have shown that the LDP algorithm can significantly improve the network capacity of URLLC (e.g., by a factor of 5-20) and can support significantly more PPRC traffic than the state-of-the-art solutions. 

Focusing on the fundamental PPRC scheduling problem for URLLC applications, this study represents a first step towards enabling URLLC in large-scale 5G-and-beyond networks with multiple channels and heterogeneous real-time requirements, 
and it serves as a foundation for exploring other interesting studies. For instance, given that the LDP scheduling algorithm and associated schedulability test are amenable to real-world implementation in cellular networks, a next-step is to implement and integrate the LDP scheduling algorithm with PRKS \cite{PRKS} in emerging open-source cellular platforms such as OpenAirInterface. 
Another interesting direction is to consider delay jitter control since URLLC applications such as XR tend to require as small delay jitter as possible.

\bibliographystyle{IEEEtran}
\bibliography{references}


{}
{
\appendices

\ifthenelse{\boolean{short}}
{}
{
\section{Proof of Theorem~\ref{NP-hard} } \label{appendix:NP-hard}

We prove the theorem by showing that the NP-hard problem of graph k-coloring can be reduced to a case of the PPRC scheduling problem in polynomial time. 

The input of k-coloring problem is as following: given a graph $G$ = ($V$, $E$) where $V$ and $E$ are the vertex set and edge set respectively. The problem is to decide whether we can color the vertices of the graph with $k$ colors such that the color of the endpoints of every edge is different. 
    
We reduce the graph k-coloring problem to the PPRC scheduling problem as follows. We assume the the first packet arrives at each link at the same time. 
For each link $i$, we also suppose that the link reliability is 1. Therefore, the required number of transmission $X_i$ equals to 1. The period $P_i$ and the relative deadline $D_i$ are both equal to $k$. The conflict graph is the same with the k-coloring problem graph $G$ = ($V$, $E$) and the number of channels is 1. This reduction can be done in polynomial time, since this reduction takes $\mathcal{O}(1)$ time for each construction.

If $G$ is k-colorable, then each vertex will be assigned a color and the endpoints of each edge will be colored differently. Then, we can construct a schedule for the links in $G$ according to the coloring: if vertex $V_i$ is colored by $s \in \{1,2,3,..,k\}$, then link $i$ can transmit its packet at the time slot $s$. Since the color of vertex $V_i$ is different with its connected vertexes, link $i$ will also transmit at different time slot compared with its one-hop neighboring links in the conflict graph. In addition, since there are totally $k$ colors for every vertex, each link in the network will complete its packet delivery in one of the $k$ time slots, that is, before the packet  deadline. Therefore, there will be no 
deadline miss in the network. 
    
Conversely, if there exists a valid schedule for the PPRC scheduling problem, we can color vertex $V_i$ in $G$ with the $s$th ($s \in \{1,2,3,..,k\}$) color if link $i$ transmits its packet at $s$th time slot. Then, since no two neighboring links are assigned with the same time slot, then no two connected vertices have the same color. This completes the proof.

\section{Proof of Theorem~\ref{thm-1} } \label{appendix:Theorem 2}
 When the iteration terminates, a link is either ACTIVE/INACTIVE based on lines~\ref{if} and \ref{done} of Algorithm~\ref{alg1}. For each INACTIVE link $i$ with non-zero local traffic demand in any channel, there always exists at least one ACTIVE link $l$, $l \in M_i$, based on lines~\ref{if inactive}, \ref{inactive} in Algorithm~\ref{alg1}. Therefore, changing any INACTIVE link to an ACTIVE link would cause two interfering links active at the same time slot in the same channel, which is not allowed. Hence, the set of all ACTIVE link for any channel is a maximal independent set.

\section{Proof of Lemma~\ref{lemma-1} } \label{appendix:Lemma1}
We can let every link $l$ in $M_{i}$ inactive and link $i$ active on all the channels, since every link $j$ in $E \setminus \{M_i\cup \{i\}\}$ does not conflict with link $i$. Then, based on the definition of minimum scheduling rate, $M_{i} \cup i$ is a  feasible set for any clique $K_{i,j}$,  $K_{i,j}\subseteq M_i \cup \{i\}$.

\section{Proof of Lemma~\ref{lemma-2} } \label{appendix:Lemma2}

We prove this by contradiction. Suppose at time slot $t$, link $i$ is not schedulable and, at time slot $t-1$, the sum of the local work density of at least one feasible set $S_{i,K_{i,j}}$ is less than $N + 1$ and suppose $X_{i,t-1} > 0$. Since $d''_{i,t-1} - (t-1) = 1$ 
at time slot $t-1$ and the work demand is an integer, $1 \leq \rho_{i,t-1} = X_{i,t-1} \leq N$. This also implies that, for the feasible set $S_{i,K_{i,j}} \subseteq M_i \cup \{i\}$, there are at most $N-X_{i,t-1}$ links whose local work density equals 1, since $\rho_{i,t-1} + \sum_{l \in S_{i,K_{i,j}}\setminus i}\rho_{l,t-1} < N + 1$. 
For each channel $rb \in RB$ and the feasible set $S_{i,K_{i,j}}$, there will be at least one active link $l \in S_{i,K_{i,j}}$. 
In addition, Algorithm 1 will let the link $l' \in M_i \cup \{i\}$ with the highest priority (whose local work density is greater than or equal to 1) be active. Therefore, each link $l \in S_{i,K_{i,j}}$ with the highest priority which is equal to 1 can be scheduled. Then, link $i$ will be active and be assigned with $X_{i,t-1}$ number of channels, and, by Definition~\ref{def:minSchRate} on Minimum Scheduling Rate, this holds no matter how the links other than those of $S_{i,K_{i,j}}$ are scheduled. 
Thus link $i$ is schedulable at time $t$, which is a contradiction. 


\section{Proof of Lemma~\ref{lemma-3} } \label{appendix:Lemma3}

According to Definition~\ref{def2} on local deadline partitioning, each link  $l \in M_i \cup \{i\}$ will choose the maximum value of the arrival time and deadline from the links in $M_l \cup l$ before time slot $t$ as $d'_{l,t}$, and choose the minimum value of the arrival time and deadline from the links in $M_l \cup l$ after time slot $t$ as $d^{''}_{l,t}$, where $d'_{l,t}$and $d^{''}_{l,t}$ are the starting time and local deadline for $\sigma_{l,t}$ respectively. This implies that, for every link $l \in M_i$ and every time slot $t$ in the period $[A_{i,p}, D_{i,p})$ associated with the $p$-th packet at link $i$, $d^{'}_{l,t} \ge A_{i,p}$ and $d^{''}_{l,t} \leq D_{i,p}$. In addition, at $t_0=A_{i,p}$, $d^{'}_{l,t_0}$ is the same for every link $l \in M_i \cup \{i\}$, and it is $A_{i,p}$; at $t_1=D_{i,p}-1$, $d^{''}_{l,t_1}$ is the same for every link $l \in M_i \cup \{i\}$, and it is $D_{i,p}$; the time slice $\left[A_{i,p}, D_{i,p}\right)$ may include multiple deadline partitions for every link $l \in M_i \cup \{i\}$.

For the feasible set $S_{i,K_{i,j}}$, since $A_{i,1}$ is the deadline of the previous deadline partition and is the same for links $l\in S_{i,K_{i,j}}\setminus {i}$, we have
\begin{equation}
\sum_{l\in S_{i,K_{i,j}}\setminus {i}}\rho_{l,A_{i,1}} + \rho_{i} \leq \sum_{l\in S_{i,K_{i,j}}}\rho_{l} = \sum_{l\in S_{i,K_{i,j}}}\frac{X_l}{D_l} \leq N.
    \label{equ5}
\end{equation}

Then we consider the total work demand (i.e., total number transmission opportunities required) for the links in $S_{i,K_{i,j}}$ during the interval $[A_{i,1},D_{i,1})$. At time slot $t_1=A_{i,1}$, every link $l \in S_{i,K_{i,j}}$ shares the same local arrival time $d_{l,t_1}'=A_{i,1}$, and, at time slot $t_2=D_{i,1}-1$, every $l \in S_{i,K_{i,j}}$ shares the same local deadline $d_{l,t_2}''=D_{i,1}$.  
Then, according to the proportionate allocation rule of the LDP scheduling algorithm (i.e., Algorithm~\ref{alg1}), during the interval $[A_{i,1},D_{i,1})$, we have 
\begin{equation}
    W_{l,A_{i,1}} = \sum_{\forall \sigma_{l,t} \subseteq [A_{i,1},D_{i,1})} \rho_{l,A_{i,1}}\times L_{l,t}, 
\end{equation}
such that $W_{l,A_{i,1}}$ is the total work demand for link $l$ in $[A_{i,1},D_{i,1})$. 


If a link $l$ has a constrained deadline (i.e., $D_i < T_i$), we have 
\begin{equation}
    \sum_{\forall \sigma_{l,t} \subseteq [A_{i,1},D_{i,1})} L_{l,t} \leq D_{i,1} - A_{i,1}. 
\end{equation}
If link $l$ has an implicit deadline (i.e., $D_i = T_i$), we have 
\begin{equation}
    \sum_{\forall \sigma_{l,t} \subseteq [A_{i,1},D_{i,1})} L_{l,t} = D_{i,1} - A_{i,1}. 
\end{equation}
Therefore, for every link $l \in S_{i,K_{i,j}}$, we have
\begin{equation}
\begin{split}
    W_{l,A_{i,1}} &= \sum_{\forall \sigma_{l,t} \subseteq [A_{i,1},D_{i,1})} \rho_{l,A_{i,1}}\times L_{l,t}\\
    &\leq (D_{i,1} - A_{i,1})\times \rho_{l,A_{i,1}}
\end{split}
\end{equation}
Then, we can get,
\begin{equation}\label{equ6}
\begin{split}
    \sum_{l\in S_{i,K_{i,j}}}W_{l,A_{i,1}} &\leq (D_{i,1}-A_{i,1}) \times \sum_{l\in S_{i,K_{i,j}}}\rho_{l,A_{i,1}}\\
    &\leq (D_{i,1}-A_{i,1}) \times N. 
\end{split}
\end{equation}

Note that meeting Condition~\ref{equ6} is critical for ensuring that link $i$ does not miss its deadline for the first packet (as it will become clear in the derivation of Conditions~\ref{equ9} and \ref{equ18} shortly), and thus it is critical for the validity of Lemma~\ref{lemma-3} too. 
    Condition~\ref{equ6} also requires the definition of deadline partitions in LDP to consider packet arrival times in addition to packet deadlines as we explain in the footnote here.\footnote{\textbf{[Including Arrival Times in Deadline Partition Definition]} If only deadlines were used in defining deadline partitions as in the traditional deadline partitioning (DP) frameworks \cite{cho2006}\cite{levin2010}, the value of $\sum_{l\in S_{i,K_{i,j}}}\rho_{l,A_{i,1}}$ at time slot $A_{i,1}$ could not be ensured to be no more than $N$, thus Condition~\ref{equ6} cannot be guaranteed to hold. 
For instance, assuming that only deadlines are used for defining deadline partitions, time $t$ and $t'$ are the beginning and end time of the current deadline partition respectively, $\sum_{l\in S_{i,K_{i,j}}}\rho_{l} = N$, and $A_{i,1} \in [t, t' )$. 
    Then, at the beginning of time slot $t$ (i.e., the beginning of a new deadline partition), $\sum_{l\in S_{i,K_{i,j}}\setminus {\{i\}}} \rho_{l,t} = \sum_{l\in S_{i,K_{i,j}} \setminus {\{i\}}} \rho_{l} $ due to proportionate traffic demand allocation and according to Definition~\ref{def4}. 
Given that the transmissions by links outside $S_{i,K_{i,j}}$ can prevent links in $S_{i,K_{i,j}}$ from transmitting at every time slot during $[t, t')$, 
$\sum_{l\in S_{i,K_{i,j}}\setminus {\{i\}}}\rho_{l,A_{i,1}} > \sum_{l\in S_{i,K_{i,j}}\setminus {\{i\}}}\rho_{l,t}$ will hold in scenarios where there is at least one link in $S_{i,K_{i,j}}$ that does not transmit in every time slot during $[t, A_{i,1})$. Then, at the beginning of time slot $A_{i,1}$, $\sum_{l\in S_{i,K_{i,j}}} \rho_{l,A_{i,1}} = \sum_{l\in S_{i,K_{i,j}}\setminus {\{i\}}}\rho_{l,A_{i,1}} + \rho_i > \sum_{l\in S_{i,K_{i,j}}\setminus {\{i\}}}\rho_{l,t} + \rho_i = \sum_{l\in S_{i,K_{i,j}}\setminus {\{i\}}}\rho_{l} + \rho_i = N$, which violates the requirement of Condition~\ref{equ6}.
}




Then we consider time slot $t_{2}=D_{i,1}-1$. Since at time $t_2$, the length of local deadline partition for all the links in $S_{i,K_{i,j}}$ is the same, the local work density can be shown as follows,
\begin{equation}
    \sum_{l\in S_{i,K_{i,j}}}\rho_{l,t_2} = \sum_{l\in S_{i,K_{i,j}}}\frac{W_{l,A_{i,1}}-C_{l,t_2}}{D_{i,1}-t_2}, 
    \label{equ7}
\end{equation}
where $C_{l,t_2}$ is the number of transmission opportunities that have been assigned to link $l$ in time slice $\left[A_{i,1}, D_{i,1}-1\right)$. We know that, 
\begin{equation}
    \begin{split}
        &\sum_{l\in S_{i,K_{i,j}}}(W_{l,A_{i,1}}-C_{l,t_2})\\
        &\leq (D_{i,1} - A_{i,1})\sum_{l\in S_{i,K_{i,j}}}\rho_{l} - \sum_{l\in S_{i,K_{i,j}}}C_{l,t_2}\\
        &\leq (D_{i,1} - A_{i,1})N - \sum_{l\in S_{i,K_{i,p}}}{C_{l,t_2}}.
    \end{split}
    \label{equ8}
\end{equation}
Based on the definition of feasible sets, we also have
\begin{equation}
\sum_{l\in S_{i,K_{i,p}}}{C_{l,t_2}} \ge N. 
\end{equation}
Thus, 
\begin{equation}
\begin{split}
    &(D_{i,1} - A_{i,1})N - \sum_{l\in S_{i,K_{i,p}}}{C_{l,t_2}} \\
    &\leq (D_{i,1} - A_{i,1})N - (D_{i,1} -1 - A_{i,1})N \\
    &\leq N.
\end{split}
\label{equ9}
\end{equation}
Therefore, based on (\ref{equ7}), (\ref{equ8}), (\ref{equ9}), we can get
{\small 
\begin{equation}
    \begin{split}
    \sum_{l\in S_{i,K_{i,j}}}\rho_{l,t_2} &= \sum_{l\in S_{i,K_{i,j}}}\frac{W_{l,A_{i,1}}-C_{l,t_2} }{D_{i,1}-t_2}\\
    & \leq \sum_{l\in S_{i,K_{i,j}}}\frac{ (D_{i,1} - A_{i,1})N - \sum_{l\in S_{i,K_{i,p}}}{C_{l,t_2}}}{D_{i,1}-t_2}\\
    & \leq \sum_{l\in S_{i,K_{i,j}}}\frac{N}{D_{i,1}-t_2}  = N 
    \end{split}
    \label{equ18}
\end{equation}
}
Therefore, according to Lemma~\ref{lemma-2}, link $i$ does not miss its deadline for the first packet, that is, the transmissions of the first packet is completed by $D_{i,1}$.
Then according to the proportionate allocation rule of the LDP algorithm, Equation~\ref{equ6} also holds for the second packet period of link $i$, $[A_{i,2},D_{i,2})$, and, based on the same analysis, this lemma holds at time slot $D_{i,2}-1$. 
By induction, the lemma also holds for any time slot $D_{i,p}-1, p \ge 3$. 
}

\ifthenelse{\boolean{short}}
{}
{
\section{Proof of Theorem~\ref{thm:sufficientCondition}}
\label{appendix:sufficientCondition}
Based on Lemma~\ref{lemma-3}, we only need to show that, for every clique $K_{i,j} \in \mathbb{K}_i$, there exists a feasible set $S_{i, K_{i,j}}$ whose sum work density is no more than $N$. This would hold if, for every clique $K_{i,j} \in \mathbb{K}_i$ and the set $\mathbb{S}_{i, K_{i,j}}$ of feasible sets for link $i$ and $K_{i,j}$, the feasible set with the minimum sum work density has a sum work density no more than $N$. Hence this theorem holds. 
}

\ifthenelse{\boolean{short}}
{}
{
\section{Proof of Theorem~\ref{thm-3} } \label{appendix:Theorem3}

According to Definition~\ref{def:feasibleSet} on feasible sets, we first need to show that there exists at least one maximal independent set (MIS) of $G_c$ whose intersection with $S_i$ has only one element. To this end, note that any MIS $mis_{G_c}$ that includes link $i$ as an element will not include any link from $M_i$, thus $mis_{G_c}$ will not include any link from $S_i \setminus \{i\}$. Therefore, for any MIS $mis_{G_c}$ such that $i \in mis_{G_c}$, $mis_{G_c} \cap S_i$ includes one and only one element $i$. 
    
Next, we need to show that there is no MIS of $G_c$ whose intersection with $S_i$ is empty. This trivially holds when $M'_i = \emptyset$. 
    When $M'_i \neq \emptyset$, for each $mis \in MIS_{M'_i}$, there must exist a MIS of $G_c$, denoted by $mis_{G_c}$, that includes $mis$ as a subset. In this case, $mis_{G_c} \cap S_i$ is not empty if and only if there is a link in $S_i$ that does not interfere with any link in $mis$. 
Of course, a MIS of $G_c$ may only include as a subset a non-maximal independent set of $M'_i$, denoted by $mis'$; in this case, there will exist a link in $S_i$ that does not interfere with any link in $mis'$, if, for for each $mis \in MIS_{M'_i}$ (which includes the $mis$ that is a superset of $mis'$), there exists at least one link in $S_i$ that does not interfere with any link in $mis$. 
    Therefore, there is no MIS of $G_c$ whose intersection with $S_i$ is empty, if and only if, for for each $mis \in MIS_{M'_i}$, there exists at least one link in $S_i$ that does not interfere with any link in $mis$.
Hence Theorem~\ref{thm-3} holds.

\section{Proof of Theorem~\ref{theorem:feasibleSetGeneration} } \label{appendix:feasibleSetGeneration}

If $S_{i,K_{i,j}}$ is a feasible set, let $M'_i = (\{i\} \cup M_i \cup M_{i,2}) \setminus S_{i,K_{i,j}}$. Then we consider $S'_{i,K_{i,j}} = S_{i,K_{i,j}} \cup K_{i,j'}$ and $M''_i = (\{i\} \cup M_i \cup M_{i,2}) \setminus S'_{i,K_{i,j}}$, and we prove that $S'_{i,K_{i,j}}$ is still a feasible set. Let $A = S'_{i,K_{i,j}}\setminus S_{i,K_{i,j}}$. 
Given the set of maximal independent sets  $MIS_{M'_i}$, there exist two sets of maximal independent sets.
Let $MIS_{M'_i,1}$ denote the set of maximal independent sets $mis_1$ such that $mis_1 \cap A \neq \emptyset $. 
Let $MIS_{M'_i,2}$ denote the set of maximal independent sets $mis_2$ such that $mis_2 \cap A = \emptyset $. 
Together, we have $MIS_{M'_i} = MIS_{M'_i,1} \cup MIS_{M'_i,2}$. Since $M''_i = M'_i \setminus A$ and the conflict graph is stable, $MIS_{M''_i} = MIS_{M'_i,2}$. Therefore, $MIS_{M''_i} \subset MIS_{M'_i}$. Then based on Theorem~\ref{thm-3}, we can know for each $mis \in MIS_{M'_i}$, there exists at least one link in $S_{i,K_{i,j}}$ that does not interfere with any link in $mis$. Since $MIS_{M''_i} \subset MIS_{M'_i}$, for each $mis' \in MIS_{M''_i}$, there exists at least one link in $S'_{i,K_{i,j}}$ that does not interfere with any link in $mis'$. Then, $S'_{i,K_{i,j}}$ is a feasible set. 

If $S_{i,K_{i,j}}$ is not a feasible set, let $M'_i = (\{i\} \cup M_i \cup M_{i,2}) \setminus S_{i,K_{i,j}}$. Then we consider $S'_{i,K_{i,j}} = S_{i,K_{i,j}} \setminus K_{i,j'}$ and $M''_i = (\{i\} \cup M_i \cup M_{i,2}) \setminus S'_{i,K_{i,j}}$, and we prove that $S'_{i,K_{i,j}}$ is still not a feasible set. Since $S_{i,K_{i,j}}$ is not a feasible set, there exists at least one $mis' \in MIS_{M'_i}$ such that any link in $S_{i,K_{i,j}}$ interfere with at least one link in $mis'$. That implies link $l' \in S_{i,K_{i,j}}$ interfere with at least one link in $mis'$. Therefore, the set of links $mis'$ is a maximal independent set for the set $M''_i$. Since each link in $S_{i,K_{i,j}}$ interfere with at least one link in $mis'$, each link in $S'_{i,K_{i,j}} \subset S_{i,K_{i,j}}$ also interfere with at least one link in $mis' \in MIS_{M''_i}$. Therefore, $S'_{i,K_{i,j}}$ is not a feasible set. 
Hence Theorem~\ref{theorem:feasibleSetGeneration} holds.
} 

\ifthenelse{\boolean{short}}
{}
{
\section{Proof of Theorem~\ref{thm-4}}
\label{appendix:Theorem6}
For any clique $K_{i,j} \in \mathbb{K}_i$, the maximum scheduling rate for each time slot is equal to the number of channels $N$. Therefore, the total utilization of the links of any clique shall be no more than $N$, where the utilization of a link $l$ is defined as $\frac{X_l}{T_l}$. Thus the theorem holds. 
}

\ifthenelse{\boolean{short}}
{}
{
\section{Proof of Theorem~\ref{Approximation ratio bound}}
\label{appendix:Theorem7}

The numerator of Equation~\ref{approximation ratio} means the maximum sum work density for all  the cliques $K_{i,j}\in \mathbb{K}_i$. To determine the lower bound of approximation ratio, let link $i$ and every other link $l \in M_i$ form a clique, which gives the minimum value of the numerator of Equation \ref{approximation ratio} (since any additional links in the clique will increase the value of the sum work density). 
    The denominator of Equation \ref{approximation ratio} means the maximum sum work density of all the feasible sets chosen by Theorem~\ref{sufficientCondition} for  schedulability test. To determine the lower bound of the approximation ratio, let all the links in $M_i \cup \{i\}$ form the chosen feasible set and then we can get the maximum value of the denominator (since removing any link from $M_i \cup \{i\}$ will decrease the value of the denominator). 
In what is next, we construct a network setting where the aforementioned properties hold. 
We let every link $l \in M_i$ and link $i$ be on the boundary of other links' exclusive regions. Then, link $i$ and every other link $l \in M_i$ form a clique of two links and the angle between every two adjacent links is exactly 60 degrees. If there were any additional interfering link $q$, it will interfere with link $i$ and any two adjacent links $l \in M_i$ and form a clique of more than 2 links, which increases the value of the numerator of Equation \ref{approximation ratio}. Therefore, the star conflict graph containing 6 links as shown in Figure~\ref{star} can determine the approximation ratio lower bound. 
    We let the sum work density of link $i$ and any link $l \in M_i$ is $\rho_0$, then we can get 
\begin{equation}
    \delta(i) = \frac{\sum_{l \in K_{i,j}}\frac{X_l}{T_l}}
    {\sum_{l \in M_i \cup i}\frac{X_l}{T_l}}
    =\frac{\rho_0}{6\frac{X_l}{T_l}+\frac{X_i}{T_i}}
    =\frac{\rho_0}{6\rho_0 - 5\frac{X_i}{T_i}} > \frac{1}{6}. 
\end{equation}
\begin{figure}[!htbp]
    \centering
    \includegraphics[width=.26\textwidth]{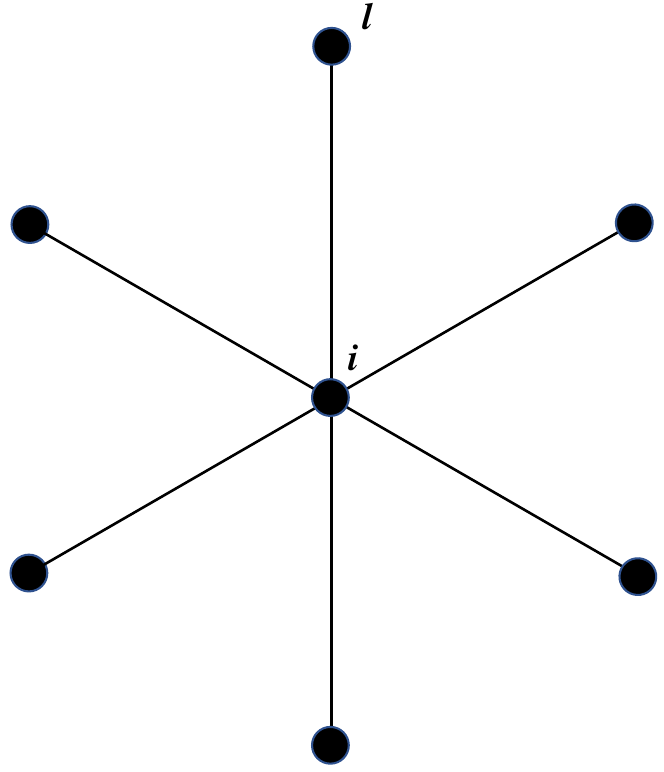}
    \caption{Star graph}
    \label{star}
\end{figure}
} }

\end{document}